\newif\ifsingle
\singletrue % comment out for single column version

\newif\ifFullVersion
\FullVersiontrue % comment out for full proofs version

\ifsingle
\documentclass[12pt,draftclsnofoot, onecolumn]{IEEEtran}		
\else		
\documentclass[10pt,final, twocolumn]{IEEEtran}
\fi

\usepackage{times}
\usepackage{amsmath,dsfont}
\usepackage{amssymb,amsthm}
\usepackage{epsfig,verbatim}
\usepackage{subcaption}
\usepackage{setspace}
\usepackage{color}
\usepackage{cite}
\usepackage{epstopdf}
\usepackage{graphics}
\usepackage{accents}
\usepackage{acronym}
\usepackage{booktabs}
\usepackage{mathtools} 
\usepackage{enumitem}
\usepackage{multirow}
\usepackage{dblfloatfix}
\usepackage[bookmarks,colorlinks]{hyperref}

\usepackage[ruled,linesnumbered]{algorithm2e}  %,algochapter

\SetKwInput{KwData}{\textbf{Init}} 
\let\oldnl\nl% Store \nl in \oldnl
\newcommand{\nonl}{\renewcommand{\nl}{\let\nl\oldnl}}% Remove line number for one line

\newcommand{\myVec}[1]{{\boldsymbol{#1}}}
\newcommand{\myMat}[1]{{\boldsymbol{#1}}}
\newcommand{\mySet}[1]{\mathcal{#1}}
\newcommand{\myY}{{\myVec{Y}}}			 		% Obsevations
\newcommand{\myS}{{\myVec{S}}}	
\newcommand{\mys}{{\myVec{s}}}	% Unknown object   
\newcommand{\myState}{\bar{\myS}}
\newcommand{\mystate}{\bar{\mys}} 
\newcommand{\myStateR}{\bar{\myVec{s}}}
\newcommand{\HyperParams}{\myVec{\theta}}
\newcommand{\Weights}{\myVec{\varphi}}

\newcommand{\Pdf}[1]{p_{ { #1}} }

\newcommand{\Mem}{L}			 			% observations length
\newcommand{\Blklen}{B}			 			% observations length
\newcommand{\Blkset}{\mySet{\Blklen}}

\ifsingle

\newcommand{\figSpace}{\vspace{-0.2cm}}

\else

\newcommand{\figSpace}{\vspace{-0.2cm}}
\fi % ------------------------------------------

\acrodef{adc}[ADC]{analog-to-digital convertor}
\acrodef{cs}[CS]{compressed sensing}
\acrodef{dtft}[DTFT]{discrete-time Fourier transform}
\acrodef{dnn}[DNN]{deep neural network} 
\acrodef{csi}[CSI]{channel state information}
\acrodef{bpsk}[BPSK]{binary phase shift keying}
\acrodef{map}[MAP]{maximum a-posteriori probability}
\acrodef{snr}[SNR]{signal-to-noise ratio}
\acrodef{bs}[BS]{base station} 
\acrodef{iot}[IOT]{Interent of Things}
\acrodef{mimo}[MIMO]{multiple-input multiple-output}
\acrodef{siso}[SISO]{single-input single-output}
\acrodef{mse}[MSE]{mean-squared error}
\acrodef{pdf}[PDF]{probability density function}
\acrodef{rv}[RV]{random variable}
\acrodef{ml}[ML]{machine learning}
\acrodef{fec}[FEC]{forward error correction}
\acrodef{rs}[RS]{Reed-Solomon}
\acrodef{lti}[LTI]{linear time-invariant}
\acrodef{wss}[WSS]{wide-sense stationary}
\acrodef{psd}[PSD]{power spectral density}
\acrodef{ser}[SER]{symbol error rate} 
\acrodef{ber}[BER]{bit error rate} 
\acrodef{gd}[GD]{gradient descent}
\acrodef{sgd}[SGD]{stochastic gradient descent} 
\acrodef{isi}[ISI]{intersymbol interference}  
\acrodef{awgn}[AWGN]{additive zero-mean white real Gaussian noise} 
\acrodef{ut}[UT]{user terminal} 
\acrodef{mmw}[mmWave]{millimeter wave}
\acrodef{noma}[NOMA]{non-orthogonal multiple access}
\acrodef{mac}[MAC]{mulitple access channel}
\acrodef{fl}[FL]{Federated learning}
\acrodef{lstm}[LSTM]{long short-term memory}
\acrodef{maml}[MAML]{model-agnostic meta-learning}
\acrodef{sic}[SIC]{soft interference cancellation}
\acrodef{pmf}[PMF]{probability mass function}
\acrodef{crc}[CRC]{cyclic redundancy check}
\acrodef{ris}[RIS]{reconfigurable intelligent surface}

\IEEEoverridecommandlockouts 

\title{Online Meta-Learning For Hybrid Model-Based Deep Receivers}
\author{  
	\IEEEauthorblockN{Tomer Raviv, Sangwoo Park,  Osvaldo Simeone, \\Yonina C. Eldar, and Nir Shlezinger
	} 
	\thanks{
		Parts of this work were %accepted for presentation 
		presented at the 2021 IEEE International Conference on Communications as the paper \cite{raviv2021meta}.	
		This project has received funding from the Israeli 5G-WIN consertium, the European Union’s Horizon 2020 research and innovation program under grants No. 646804-ERC-COG-BNYQ and No. 725731. 
        Support is also acknowledged from the Israel Science Foundation under grant No. 0100101.
		T. Raviv and N. Shlezinger are with the School of ECE, Ben-Gurion University of the Negev, Beer-Sheva, Israel (e-mail: tomerraviv95@gmail.com; nirshl@bgu.ac.il).  
		S. Park and O. Simeone are with the Department of Engineering, King’s College London,  U.K. (email: \{sangwoo.park; osvaldo.simeone\}@kcl.ac.uk).
		Y. C. Eldar is with the Faculty of Math and CS, Weizmann Institute of Science, Rehovot, Israel (e-mail: yonina.eldar@weizmann.ac.il).}

	\vspace{-1.0cm}
	
}
\vspace{-0.75cm}

\begin{document}
	
	\maketitle

	\pagestyle{plain}
	\thispagestyle{plain}
	%----------------------------------------------------------------------------------------
	%	ABSTRACT
	%----------------------------------------------------------------------------------------
	\begin{abstract} 
Recent years have witnessed growing interest in the application of deep neural networks (DNNs) for receiver design, which can potentially be applied in complex environments without relying on knowledge of the channel model. However, the dynamic nature of communication channels often leads to rapid distribution shifts, which may require periodically retraining. 
This paper formulates a data-efficient two-stage training method that facilitates rapid online adaptation. Our training mechanism uses a predictive meta-learning scheme to train rapidly from data corresponding to both current and past channel realizations. Our method is applicable to any \ac{dnn}-based receiver, and does not require transmission of new pilot data for training. To illustrate the proposed approach, we study \ac{dnn}-aided receivers that utilize an interpretable model-based architecture, and introduce a modular training strategy based on predictive meta-learning. 
We demonstrate our techniques in simulations on a synthetic linear channel, a synthetic non-linear channel, and a COST 2100 channel. Our results demonstrate that the proposed online training scheme allows receivers to  outperform previous techniques based on self-supervision and joint-learning by a margin of up to 2.5 dB in coded bit error rate in rapidly-varying scenarios.
\end{abstract}
	%----------------------------------------------------------------------------------------
	%	Introduction
	%----------------------------------------------------------------------------------------
	%\vspace{-0.4cm}
	\section{Introduction}
\label{sec:introduction}
\vspace{-0.1cm} 
	
% Paragraph - need to operate in complex channel conditions
% \RevisionHighlights{Traditional receiver design is dominated by methods that are based on statistical models. Such conventional receivers rely on  mathematical models that describe the transmission process, signal propagation, interference, and many other components of the system that affect end-to-end communications. These statistical models typically adopt simplifying assumptions 
% that make them tractable and understandable, and use parameters that vary over time as the channel conditions change.   The need for an accurate and simplified model may constitute a notable limitation in future communication technologies.  As high frequency bands and spectrum sharing  become widespread, current simplified models used for the channel, interference, and noise may be inaccurate \cite{xia2020mimo}. Moreover, emerging antenna architectures result in complex modelling of the transmission and reception process \cite{shlezinger2020dynamic}, while the deployment of intelligent surfaces notably complicates channel estimation \cite{alexandropoulos2021hybrid}. Furthermore, signalling in high frequencies is often prone to hardware non-idealities affecting the overall model, either unintentional or due to using limited hardware, e.g., few-bit \acp{adc} \cite{shlezinger2018asymptotic} and non-linear amplifiers \cite{singya17PA}.  This makes conventional channel-model-based techniques significantly more complex.}
	
% Paragraph - deep learning for receivers
Deep learning systems have demonstrated unprecedented success in various applications, ranging from computer vision to natural language processing, and recently also in physical layer applications. 
While traditional receiver algorithms are channel-model-based, relying on mathematical modeling \cite{xia2020mimo} of the signal transmission, propagation, and reception. \ac{dnn}-based receivers have the potential to operate efficiently in model-deficient scenarios where the  channel model is unknown, highly complex \cite{shlezinger2020dynamic,alexandropoulos2021hybrid}, or difficult to optimize for~\cite{farsad2018neural,shlezinger2018asymptotic}. Generally, deep learning can be integrated with receiver design either by using conventional black-box DNN architectures trained end-to-end; or by leveraging model-based solutions \cite{shlezinger2022model,shlezinger2020model,eldar2022machine,cammerer2020trainable}, whereby specific blocks of a receiver's architecture are replaced  by neural networks, e.g., via deep unfolding \cite{shlezinger2020model}. Contrary to black-box receivers, which make limited assumptions on the data distribution, model-based deep receivers exploit additional domain knowledge, in the form of a specific receiver structure that is tailored to the channel of interest.

Channel encoding and decoding can be optimized jointly end-to-end as in \cite{ye2018channel,ye2021deep,farsad2018deep}, or decoding can be separately trained as studied in \cite{nachmani2018deep,gruber2017deep,be2019active,nachmani2019hyper}. Classical channel estimation can be enhanced as compared to the existing compressive sensing-based methods \cite{he2018deep,wen2018deep} via the integration with \ac{dnn}s. \ac{dnn}-aided receiver designs are shown to outperform classical methods in non-linear environments in \cite{gunduz2019machine,simeone2018very,balatsoukas2019deep, oshea2017introduction, zheng2021deep}. In the areas of optical fiber communication and underwater acoustics, in which non-linearity dominates, \ac{dnn}-aided solutions have also proved useful \cite{karanov2018end,zhu2019autoencoder,uhlemann2020deep,lu2021deep,zhang2022deep}. Additional  applications of \ac{dnn}-aided receivers include detection by \ac{ris} and blind reception with multiple modulation and coding schemes \cite{khan2019deep,xu2021artificial}.

% Paragraph - inherent challenge due to temporal variations
Despite their potential in implementing digital receivers \cite{shlezinger2020inference,farsad2020data}, deep learning solutions are subject to several challenges that limit their applicability in important communication scenarios. A fundamental difference between digital communications and traditional deep learning applications stems from the dynamic nature of wireless channels.  \acp{dnn} consist of highly-parameterized models that can represent a broad range of mappings. As such, massive data	sets are typically required to obtain a desirable mapping. 
The dynamic nature of communication channels implies that a \ac{dnn} trained for a given channel may no longer perform well on future channel realizations. While one can possibly enrich data sets via data augmentation~\cite{kim2022massive,raviv2022adaptive,soltani2020more} or robustify a \ac{dnn} so that it copes with multiple channel realizations via Bayesian learning~\cite{zecchin2022robust,nikoloska2022bayesian}, a \ac{dnn}-aided receiver is still likely to have to adapt at some point when operating in time-varying conditions.

%This paper addresses this problem via online meta-learning and hybrid model-based deep receivers. Specifically, we optimize the online training algorithm to incorporate previous knowledge. We also refer to other papers that address the train/test distributions mismatch that arises in time-varying wireless communication channels by means of either data augmentation \cite{kim2022massive,zhang2022data,raviv2022adaptive,soltani2020more,rizk2019effectiveness,erpek2020deep}, or by Bayesian learning \cite{zecchin2022robust,nikoloska2022bayesian}.

To apply \ac{dnn}-based transceivers in time-varying channels, two main approaches are considered in the literature. The first  attempts to learn a single mapping that is applicable to a broad range of channel conditions. This class of methods includes the approach of training a \ac{dnn} using data corresponding to an extensive set of expected channel conditions, which is referred to as {\em joint learning}\cite{oshea2017introduction, xia2020note}. An additional, related, method trains in advance a different network for each expected channel, and combines them as a deep ensemble \cite{raviv2020data}. However, these strategies typically require a large amounts of training data, and  deviating from the training setup, i.e., operating in a channel whose characteristics differ from those observed during training, can greatly impair performance \cite{park2020learning}. 

An alternative strategy is to track the channel variations. This can be achieved by providing the \ac{dnn} with a model-based channel estimate \cite{he2020model,honkala2021deeprx,khani2020adaptive, samuel2019learning, goutay2021machine}. However,  channel estimation involves imposing a relatively simple model on the channel, such as a linear Gaussian models, which may be inaccurate in some setups. 
When operating without channel knowledge, tracking the channel involves periodically retraining the network. To provide data for retraining, one must either transmit frequent pilots, or, alternatively, use decoded data for training with  some \ac{fec} scheme. 
Specifically, the mechanism employed in  \cite{shlezinger2019viterbinet, teng2020syndrome,schibisch2018online} re-encodes the decoded bits, and then it computes the Hamming distance between the re-encoded bits and the hard-decision obtained from the channel observations. If this difference (normalized to the block length) is smaller than some threshold value, the decoded bits are considered to be reliable and are used for retraining. Another approach is to further encode the block with error detection codes such as \ac{crc}, which identifies decoding errors. If the received block is deemed to be detected correctly, as indicated by the above measures, it is fed back into the \ac{dnn}-aided receiver, with the predicted symbols as the training labels. This process implements a form of self-supervised training as defined in \cite{mao2020survey,jaiswal2020survey,wang2022self}.

Data-driven implementations of the Viterbi scheme \cite{viterbi1967error},  BCJR method \cite{bahl1974optimal}, and  iterative \ac{sic}  \cite{choi2000iterative} were proposed in \cite{shlezinger2019viterbinet,shlezinger2020data, shlezinger2019deepSIC}, respectively. Yet, even with model-based architectures, relatively large data sets are still required and typical \ac{dnn} training procedures are likely to induce non-negligible delay. The ability to retrain quickly is highly dependent on the selection of a suitable initialization of the iterative training algorithm. While the common strategy is to use random weights, the work \cite{shlezinger2019viterbinet} used the previous learned weights as an initial point for  retraining. 
An alternative approach is to optimize the initial point via meta-learning \cite{park2020meta,simeone2020learning,jiang2019mind,park2020learning,park2020end}. Following this method, one not only retrains, but it also optimizes the hyperparameters that define the retraining process.   
Meta-learning was adopted to facilitate prediction of blockages~\cite{kalor2021prediction}, beam tracking~\cite{yuan2020transfer}, and power control~\cite{nikoloska2021fast}.

For \ac{dnn}-aided receivers, it was shown in \cite{park2020learning} that by optimizing the initial weights via meta-learning, the receiver can quickly adapt to varying channel conditions. The method proposed in \cite{park2020learning} is designed for settings where in each coherence duration, the transmitters send pilots followed by data packets, such that the receiver can train using the pilots and then apply its \ac{dnn} to the received data. This technique does not naturally extend to rapidly time-varying channels. In fact, in such cases, the channel can change between the  packets and thus a receiver trained with the pilots may no longer be suitable for detecting the following information messages.

In this work we propose an online training algorithm to enable rapid adaptation of \ac{dnn}-based receivers via meta-learning. Our algorithm considers both long and short term variations in the channel: We choose the initial weights of the deep receiver via meta-learning \cite{finn2017model}, while tracking local variations in a decision-directed self-supervised manner
%. The latter is achieved by exploiting the presence of \ac{fec} codes, commonly utilized in digital communications, \SangwooCmt{to retrain by detection, using  re-encoded decoded blocks as training data
\cite{shlezinger2019viterbinet,teng2020syndrome}. 
While meta-learning in \cite{park2020learning} is designed to train from pilots corresponding to the same channel over which the information blocks are transmitted, we consider rapidly time-varying channels, where each block undergoes a different channel realization. Consequently, while \cite{park2020learning} was able to build upon the conventional \ac{maml} algorithm \cite{finn2017model}, our proposed approach modifies \ac{maml} to incorporate predictions of channel realizations in future blocks. %Usages of meta-learning in communication, other than in receivers, include predicting blockages in links over high carrier frequencies \cite{kalor2021prediction}, and fast adaptation for beamforming \cite{yuan2020transfer} or power control \cite{nikoloska2021fast}. However, not many works employ this algorithm for the online detection task in time-varying channels and tailor it to this specific usecase as we propose.
%which considered online training with pilots,  while we consider training with data blocks unknown to the receiver. Moreover, we distill predictive abilities into the weights by incorporating the subsequent transmission of blocks into the training loss, which \cite{park2020learning} did not consider.}

% TODO NIR - continue from here

To further facilitate efficient adaptation to changing conditions, we instantiate the proposed meta-learning approach for settings in which the \ac{dnn}-aided receiver employs an interpretable model-based deep architecture \cite{shlezinger2020model}. Hybrid design techniques, such as deep unfolding \cite{monga2021algorithm}, neural augmentation \cite{satorras2020neural}, and \ac{dnn}-based algorithms \cite{farsad2021data}, have given rise to a multitude of deep receiver architectures, including, e.g., \cite{shlezinger2020data,samuel2019learning,he2020model,nachmani2018deep,balatsoukas2019deep, be2019active,shlezinger2019deepSIC, shlezinger2019viterbinet}. These hybrid  model-based/data-driven deep receivers are generally more compact in terms of number of parameters, requiring fewer training samples to reach convergence \cite{shlezinger2020model}.
We propose to exploit the structure of model-based deep receivers in order to efficiently re-train only specific modules of the architecture with the aid of meta-learning \cite{raghu2019rapid}. This new approach to online training, referred to as \textit{modular training},  further reduces the overall error rate over rapidly-changing scenarios.

% to train This enabling the use of self-generated labels that extend the availability of supervised data beyond the pilot blocks. 
Our main contributions are summarized as follows:
\begin{itemize}
    \item \textbf{Predictive online meta-learning algorithm:} We introduce a meta-learning strategy that is applicable to any DNN-based receiver. The proposed approach integrates \ac{maml} \cite{finn2017model} with the prediction of time-varying channels to ensure that a small number of gradient updates can efficiently minimize the loss (or error) on the next data block. %The method developed is generic, and applicable to any \ac{dnn}-based receiver without inducing any pilot overhead at test time.
    \item \textbf{Modular training of interpretable architectures:} 
    We instantiate the proposed meta-learning approach for hybrid model-based/data-driven deep receivers that leverage the structure of multipath channels. We propose to only retrain specific modules of the receiver architecture that need adaptation due to temporal variations of the channel. Fast and efficient adaptation of the machine learning modules utilize a variant of MAML referred to as almost-no-inner-loop  (ANIL) \cite{raghu2019rapid}, which incorporates \emph{a priori} information about channel variability.
    \item \textbf{Extensive Experimentation:} We extensively evaluate the proposed training scheme for both \ac{siso} and \ac{mimo} systems, considering three different time-varying channel profiles: linear synthetic channel, a non-linear synthetic channel, and a COST 2100 \cite{liu2012cost}  channel.
    We show gains of up to 2.5 dB in coded \ac{ber}, compared to joint and online approaches, over a set of challenging channels. Compared to previous approaches for handling time-varying channels without relying on pilots, such as self-supervision \cite{shlezinger2019viterbinet} and joint learning \cite{xia2020note}, we show that the proposed techniques offer advantages in terms of average \ac{ser}, while reducing the pilot overhead.
\end{itemize}

 The rest of this paper is organized as follows:  Section~\ref{sec:Model} details the system model, while Section~\ref{sec:Hybrid Meta-Learning for Data-Driven Receivers} presents the meta-learning algorithm. In Section~\ref{sec:modular-training} we focus on model-based deep receivers and show how to exploit their interpretable structure via modular training. Finally, experimental results and concluding remarks are detailed in Section~\ref{sec:Simulation} and Section~\ref{sec:Conclusions}, respectively.

Throughout the paper, we use boldface letters for vectors, e.g., ${\myVec{x}}$; $({\myVec{x}})_i$ denotes
the $i$th element of ${\myVec{x}}$. %We use upper-case letters for \acp{rv}, and lower-case letters for deterministic quantities. 
Calligraphic letters, such as $\mySet{X}$, are used for sets, 
and $\mySet{R}$ is the set of real numbers.

	%----------------------------------------------------------------------------------------
	%	System Model
	%----------------------------------------------------------------------------------------
	%\vspace{-0.2cm}
	\section{System Model}
\label{sec:Model}
\vspace{-0.1cm}

In the section, we describe the system model in Subsection~\ref{subsec:Channel}, and then formulate the design problem in Subsection~\ref{subsec:Problem}. 

\begin{figure*}[!t]
    \centering
    \includegraphics[width=\textwidth,height=0.25\textheight]{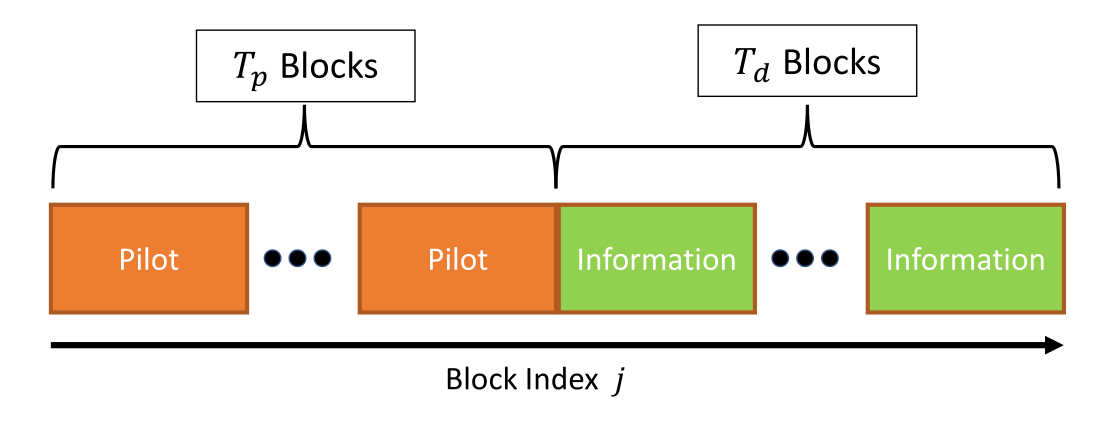}
    \caption{Transmission model. The channel is constant in each block and  changes across blocks. }
    \figSpace
    \label{fig:transmission}
\end{figure*}

\vspace{-0.2cm}
\subsection{Channel Model}
\label{subsec:Channel}
\vspace{-0.1cm} 

We consider communication over casual finite-memory block-wise varying channels with multiple users. The channel output depends on the last $\Mem >0$ transmitted symbols, where $\Mem$ is the memory length. The channel is constant within a block of $\Blklen$ channel uses, which corresponds to the coherence duration of the channel. 

Let $\myVec{S}_{i,j}\in\mySet{S}^K$ be the $K$ symbols transmitted from constellation $\mySet{S}$ at the $i$th time instance, $i \in \{1,2,\ldots, \Blklen\}:= \Blkset$, of the $j$th block. Here, the constellation size is $|\mySet{S}|$, and $K$ denotes the number of users transmitting simultaneously. Accordingly, when $K>1$, the setting represents a multiple access channel, while $K=1$ corresponds to point-to-point communications.
The channel output is denoted by $\myY_{i,j}\in \mySet{Y}^N$, where $N$ represents the number of receive antennas, and is given by a stochastic function of the last $L$ transmitted symbols 
%$\myVec{S}_{i,j-\Mem+1},\ldots, \myVec{S}_{i,j}$.
$\myState_{i,j} := [\myVec{S}_{i-\Mem+1,j},\ldots, \myVec{S}_{i,j}]^T$.  
Specifically, by defining the $j$th transmitted block as $\myS_j := \{\myVec{S}_{i,j}\}_{i\in \Blkset}$ and its corresponding observations as $\myY_j:= \{\myY_{i,j}\}_{i\in \Blkset}$,  the conditional distribution of the channel output given its input satisfies   
\begin{equation}
	\label{eqn:ChModel1}
	P_{\myY_j | \myS_j}\left(\myVec{y}_j | \myVec{s}_j \right)  = 
	\prod\limits_{i\!=\!1}^{\Blklen}P_{\myY_{i,j} |\myState_{i,j}}\left(\myVec{y}_{i,j} |\mystate_{i,j}\right).  
\end{equation}
In \eqref{eqn:ChModel1}, the lower-case notations $\myVec{s}_j$, $\myVec{y}_{i,j}$, and $\myStateR_{i,j}$ represent the realizations of the \acp{rv} $\myS_j$, $\myY_{i,j}$, and $\myState_{i,j}$, respectively. We set $\myVec{S}_{i,j} \equiv \myVec{0}$ for $i<0$, i.e., we assume a guard interval of at least $\Mem-1$ time instances between blocks.

%Practical communication scenarios are highly complex; 

The input-output relationship in \eqref{eqn:ChModel1} describes a generic, possibly multi-user, channel model with block-wise temporal variations. These  channel variations, i.e., changes between different values of the block index $j$, can often be attributed to some phenomena, e.g. a  movement  of one or several of the communicating entities. %Our two-stage training method is tailored for these complex scenarios, exploiting domain knowledge, whereas naive algorithms fail. 
We do not impose a specific model on the channel observed in each block, representing it by the generic conditional distribution in \eqref{eqn:ChModel1}, which can take a complex and possibly intractable form. Two important special cases include:
\begin{enumerate}
    \item \emph{\ac{siso} Finite-Memory Channels}: The first type of channels considered in this work is the multipath \ac{siso} case; that is, we set $N=K=1$. 
    \item \emph{Flat \ac{mimo} Channels}: Another channel of interest is the memoryless ($L=1$) \ac{mimo} channel, where $N>1$ and $K>1$. Such settings specialize multi-user uplink \ac{mimo} systems, where $K$ is the number of single-antenna transmitters. 
\end{enumerate}

We consider the scenario illustrated in Fig.~\ref{fig:transmission}, where a total of $T_p$ pilot blocks and $T_d$ information blocks are transmitted sequentially. That is, blocks indexed $j\in\{0,\ldots,T_p-1\}$ are known pilots, transmitted in the training phase,  while $j\in\{T_p,\ldots,T_p+T_d-1\}$ are information blocks that compose the test phase. The pilot blocks convey a known message, whereas the information blocks are unknown and are utilized to test the proposed schemes. Each information block, i.e., $\myVec{S}_j$ for $j\in\{T_p,\ldots,T_p+T_d-1\}$, is encoded into a total of $B$ symbols using both \ac{fec} coding and error detection codes. Error detection codes, such as cyclic redundancy check, allow the receiver to identify the presence of errors in the decoding procedure.

%-----------------------------------
%	Problem Formulation
%-----------------------------------
\vspace{-0.2cm}
\subsection{Problem Formulation}
\label{subsec:Problem}
\vspace{-0.1cm}  
We consider the problem of symbol detection in the generic channel model formulated in Subsection~\ref{subsec:Channel}. The fact that we do not impose a specific model of the channel, allowing its input-output relationship to take complex forms in each block, combined with the availability of data based on the protocol detailed above, motivate using \ac{dnn}-based receivers. The core challenge in applying \ac{dnn}-based receivers for the considered channel stems from its temporal variations. Our objective is thus to derive a training algorithm that aids \ac{dnn}-based receivers in recovering the transmitted data from the channel outputs  $\{\myVec{Y}_j\}_{j=T_p}^{T_p+T_d-1}$. The \ac{dnn}-based receiver trains on the pilot blocks, and is tested on the transmitted information blocks.

Let $\Weights$ denote the parameters of the \ac{dnn}-based receiver, i.e., the weights of the neural network. These parameters dictate the receiver mapping, which, for a given $\Weights$, is denoted by $\hat{\myVec{s}}(\cdot; \Weights):\mySet{Y}^{N\times \Blklen } \mapsto \mySet{S}^{K\times \Blklen}$, i.e., $\hat{\myVec{s}}(\myVec{y}; \Weights)$ denotes the symbol block recovered\footnote{while the \ac{dnn}-based receiver is formulated here as outputting symbol decisions, one can also adapt the formulation to soft (probabilistic) decisions. } by the receiver parameterized with $\Weights$ when applied to the channel output block $\myVec{y}$. By using the abbreviated symbol  $\hat{\myVec{s}}_j(\Weights)$ to denote the symbols detected based on the channel output of the $j$th block, with $\hat{\myVec{s}}_{i,j}(\Weights) \in \mySet{S}^K$ being its $i$th symbol, $i \in \Blkset$, our goal is to propose an algorithm that allows a \ac{dnn}-based receiver  to  continuously provide low error rates over the data blocks:
\begin{equation}
\label{eqn:ErrorRate}
	\mathop{\min}_{\Weights_j}\Big(\frac{1}{\Blklen}\sum_{i=1}^{\Blklen} \Pr\left( \hat{\myVec{s}}_{i,j}(\Weights_j) \neq \myVec{s}_{i,j} \right)\Big), \qquad j\in\{T_p,\ldots,T_p+T_d-1\}. 
\end{equation}
The parameter vector $\Weights$ in \eqref{eqn:ErrorRate} is allowed to change between blocks, namely, the \ac{dnn}-based receiver can adapt its parameters over time, while the pilot blocks are utilized to initialize $\Weights$.

As our problem formulation does not impose a specific structure on the \ac{dnn}-based receiver, we first consider  online adaptation that is agnostic of the receiver structure. This approach integrates channel predictions into the adaptation of the receiver's parameters. Next, we show how one can employ the modular architecture of model-based deep receivers to further reduce the error rate over the transmitted data blocks, facilitating coping with rapid variations and a short coherence duration.

	%----------------------------------------------------------------------------------------
	%	Meta-Learning Method
	%----------------------------------------------------------------------------------------
	%\vspace{-0.2cm}
	\section{Predictive Online Meta-Learning for DNN-Based Receivers}
	\label{sec:Hybrid Meta-Learning for Data-Driven Receivers}
	\vspace{-0.1cm}
	In this section we study parameter adaptation of a generic \ac{dnn}-based receiver to continuously maintain low error rates in time-varying channels. Modification of these weights involves training, which requires the receiver to obtain recent examples of pairs of transmitted blocks and corresponding observed channel outputs. This gives rise to two core challenges. First, while one can use pilots, the time-varying nature of the channel implies that  pilot-based examples are likely to not represent the  channel conditions for some or all of the data blocks. In addition, even if one can extract labeled training data from blocks containing information messages, when the channel varies quickly, the amount of data from the instantaneous channel may be limited. We draw inspiration from the successful applications of meta-learning for facilitating training from pilots corresponding to the current channel conditions \cite{park2020learning}, as well as the emergence of \ac{fec}-based self-supervision for pilot-free online training  \cite{shlezinger2019viterbinet}. 
	%We fuse these two concepts together into a time-predictive online scheme which leverages both past decoded blocks (as in meta-learning) as well as the current ones (as in \ac{fec}-based online training) to facilitate rapid retraining and coping with temporal variations.}% Combining this with the typical lengthy duration of \ac{dnn} training procedures makes online adaptation of \ac{dnn}-based receivers a challenging task. 
	
	%These challenges are tackled by the proposed online meta-training algorithm. 
	To describe this algorithm, we begin by introducing self-supervised training in Subsection~\ref{subsec:Online} for slow-fading channels. Then we extend online training to rapidly time-varying channels in Subsection~\ref{subsec:high-level}, which presents our training method. The resulting algorithm is detailed in Subsection~\ref{subsec:MetaLearningDiscussion}.

%-----------------------------------
%	Decision-Directed Online Training
%-----------------------------------
\vspace{-0.2cm}
\subsection{Self-Supervised Online Training}
\label{subsec:Online}
\vspace{-0.1cm} 

% \textcolor{red}{At this point in the paper we have already formulated the channel model and the problem. So now we need to gradually lead towards our solution based on the model. So I recommend starting the section by:\\
% Start by explaining that since we consider the operation of \ac{dnn}-based receivers in complex channels, as detailed in Subsection~\ref{subsec:Channel}, one would need to periodically re-train the \ac{dnn} in order to preserve reliable operation in time-varying conditions. \\
% Then, explain that the core algorithmic challenge in doing so stems from the availability of data. Training requires data corresponding to the instantaneous channel conditions being observed currently, while the generic protocol detailed in Subsection-??? implies that pilots, which can be used to formulate a labeled training set, are scarce and may not correspond to the current channel. }

% We regard the problem of \ac{dnn}-based receivers in complex time-varying channels, as detailed in Subsection~\ref{subsec:Channel}, where the distribution of the channel output $\myY$ changes with time. As the deviation between the trained-on channel and the tested one increases, one obviously needs to periodically re-train the receiver to preserve high performance operation. 

The first part of our online adaptation mechanism is based on self-supervision, which extracts training data from information-bearing blocks. The approach follows \cite{shlezinger2019viterbinet} by re-using confident decisions via the re-encoding of a successfully decoded word for use in online training. 
%
%Following this approach, the parameters of a \ac{dnn}-based receiver are adapted on a block by block basis; That is, on any incoming block $\myVec{y}_j$. Even if the deep symbol detector is not trained for the instantaneous channel realization from which $\myVec{y}_j$ is generated and thus its estimate $\hat{\myVec{s}}_j$ may differ from the true transmitted symbol vector, the \ac{fec} decoder can still correct these errors, as long as they are within the Hamming distance of the code. Specifically, the channel decoder takes as input the estimated block $\hat{\myVec{s}}_j$, and outputs a decoded message along with the number of errors, as an indication of the correctness of its decoded message. When decoding is correct, i.e. the number of errors is 0, the decoded and demodulated message $\hat{\myVec{m}}_j$ is encoded and modulated into the estimated transmitted symbols  $\myVec{s}_j$.

\begin{figure}
    \centering
    \includegraphics[width=0.9\columnwidth]{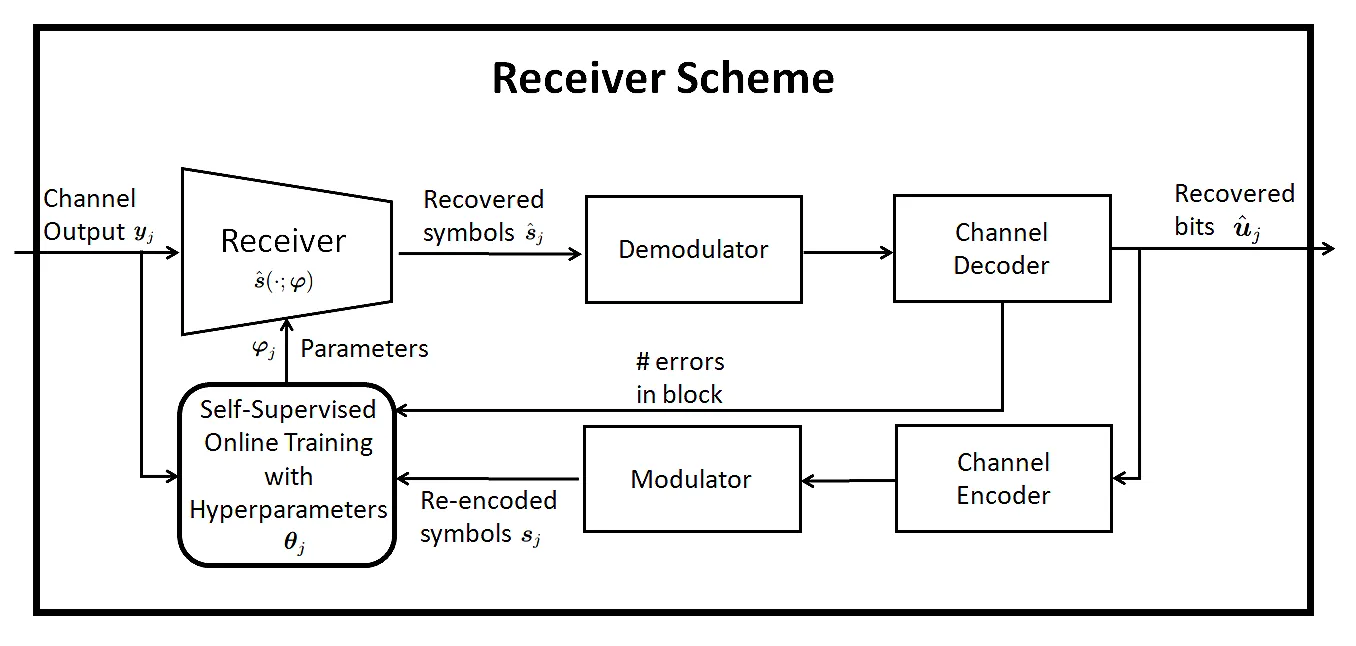}
    \vspace{-0.6cm}
    \caption{Self-supervised online training based on channel coding.}
    \label{fig:self-supervised training}
\end{figure}

% \textcolor{red}{We can consider adding a figure visualizing how online training operates. It can be divided into two parts - symbol detection and online training. }

% Specifically, the detector for the $j$th block is parameterized by the weight vector $\Weights_j$, which is applied to $\myVec{y}_j$. When decoding is successful, one can confidently generate a 

First, we initialize the parameter vector $\Weights_{T_p}$ by training over the obtained $T_p$ pilot blocks. Then, for each information block $j\in\{T_p,\ldots,T_p+T_d-1\}$, when decoding is reported by the decoder to be correct, the re-encoded data pair $\{\myVec{s}_{j}, \myVec{y}_{j}\}$ is used as training data to update the parameters for the next block, i.e., to generate $\Weights_{j+1}$. The same training is applied to the receiver during pilots and information blocks transmission. Since symbol detection can be treated a classification task, this is achieved by minimizing the empirical cross entropy loss as
\begin{align}
\label{eq:meta_1}
  \mathop{\arg \min}_{\Weights} \left\{ \mySet{L}_{j}(\Weights) = -\sum_{i=1}^{\Blklen}\log \hat{P}_{\Weights}\left(\myVec{y}_{i,j}|\myStateR_{i,j} \right)\right\}.
\end{align}
In (\ref{eq:meta_1}), the  probability $\hat{P}_{\Weights}\left( \myVec{y}  |  \myStateR\right)$ is the soft output of the \ac{dnn}-based receiver corresponding to $\myStateR$ when applied to the data block $ \myVec{y}$. The optimization problem in \eqref{eq:meta_1} is approximately solved using gradient-based optimization with some learning rate $\eta$, i.e., via $I_{\rm sgd}$ iterations of the form
 \begin{align}
\label{eq:local_update} % \label{eq:meta_1}
\Weights_{j+1}^{(t+1)} = \Weights_{j+1}^{(t)} - \eta \nabla_{\Weights_{j+1}^{(t)}}\mySet{L}_{j}(\Weights_{j+1}^{(t)}).
 \end{align}
 %where $\eta > 0$ is the learning rate. 
 
 The computation of the gradient steps requires an initialization $\Weights_{j+1}^{(0)}$, which is a hyperparameter of the training procedure. We henceforth denote this hyperparameter vector used for self-supervised training of $\Weights_{j+1}$ as  $\HyperParams_{j+1}$, i.e., $\Weights_{j+1}^{(0)} = \HyperParams_{j+1}$. 
 The gradient in \eqref{eq:local_update} is typically approximated via random sampling among available $B$ symbols within the block to implement \ac{sgd} and its variants, typically used for training DNNs \cite[Ch. 4]{goodfellow2016deep}.  The online training scheme is presented in Fig.~\ref{fig:self-supervised training}.
%
%In order to enable an adaptation mechanism, the receiver maintains at each block index $j$ a vector of hyperparameters $\HyperParams_j$. %\NirCmt{Do we need to introduce the data buffer at this stage? We do not really use it here, but for meta-learning. So I guess the main question is whether the purpose of this section is to discuss how online self-supervised training is done in the literature, or to explain how we do it here? If we go with the latter, we need to explain that the proposed scheme only observes short-term variations, i.e., those between the current block and the next one, and thus missing out on long-term variation, such as patterns of variations observed in the past. To be able to also account for these longer variations, we save decisions in the buffer, whose usage is revealed in ths sequel.}.
%
%At each data block $j$, given the initialization hyperparameter vector\NirCmt{Perhaps we should not state here how the initial hyperparameters vector is set, just say that we use some initialization $\HyperParams_j$. Then, when we conclude the section say that the approach used in \cite{shlezinger2019viterbinet} used $\HyperParams_j=\Weights_{j-1}$, which implies that only short-term variations are accounted for in the training procedure, as both the data and the optimization hyperparameters correspond to the current instantaneous channel realization.} $\HyperParams_j$ and successfully decoded blocks' pair $\{\myVec{s}_{j-1}, \myVec{y}_{j-1}\}$, the algorithm updates the model parameters vector $\Weights_j$. 
 %
 The self-supervision training algorithm is based on the assumption that channel conditions vary smoothly across blocks, so that a receiver trained on data from the $j$th block is likely to correctly detect data also from the $(j + 1)$th channel realization. Following this line of reasoning,  \cite{shlezinger2019viterbinet} used $\HyperParams_{j+1}=\Weights_{j}$. 
 
 %two main characteristics of digital communications over time-varying channels: 
  %\begin{enumerate}
    %\item The first is the presence of \ac{fec} coding that provides a safe margin and allowing one to recover the transmitted symbols even as a small number of errors occur. The exact number is dependent on the Hamming distance of the \ac{fec} code. Moreover, one can identify when re-encoding is possible, employing the re-encoded block for training. 
    
    %\item The second property exploited is the fact that dynamic channels are of continuous nature; Thus one can reasonably assume that the data from the channel observed in the $j$th time index is still applicable to the $j+1$th channel realization. Following this line of reasoning, the researchers in  \cite{shlezinger2019viterbinet} used $\HyperParams_{j+1}=\Weights_{j}$. 
 %\end{enumerate}
 
However, this assumption is inadequate for tracking fast-varying channels. Adapting locally in a self-supervised manner by setting the hyperparameter $\HyperParams_{j+1}=\Weights_{j}$ accounts for only short term variations, as both the data and the optimization hyperparameters correspond to the current instantaneous channel realization. In order to train effectively from short blocks in complex channels, we propose to incorporate long term relations by altering the setting of the hyperparameter $\HyperParams_{j+1}$ via meta-learning.
 
 %mini-batch. Note  \eqref{eq:local_update} describes a single \ac{sgd} iteration,  multiple iterations are similarly accommodated.

% \textcolor{red}{I recommend concluding the above paragraph by clarifying that the ability to carry out such self-supervision relies on two characteristics of digital communications over time-varying channels: The first is the presence of coding, which allows one to recover re-form the transmitted symbols even when detection has errors as long as these errors are below the Hamming distance of the \ac{fec} code, as well as identify when such re-encoding is possible due to error detection codes. The second property exploited is the fact that dynamic channels are not arbitrarily time-varying, and thus one can train using data corresponding the channel observed during the $j$th while still being applicable in the $j+1$th channel realization. \\
% After saying that, please explain why is this not enough on its own. This will lead us to the need optimize the hyperparameters $\theta_j$ in order to effective online training from short blocks and with few epochs, which is discussed in the following section. This way you lead your readers through your method (which on its own is quite complicated, thus requires a gradual gentle well-motivated introduction of its components). }

%-----------------------------------
%	High-Level Description
%-----------------------------------
\vspace{-0.2cm}
\subsection{Predictive Online Meta-Learning} 
\label{subsec:high-level}
\vspace{-0.1cm}
% While online training as described in the previous subsection is successful when the channel variations are relatively slow, i.e., when the block size $\Blklen$ is large enough to allow the aggregation of sufficient self-supervised data, it may fail to track fast changing channels. To tackle this problem, we propose to exploit long-term memory based on previously observed channel variation patterns using meta-learning to optimize the hyperparameter $\HyperParams_{j}$.

%, in contrast to the self-supervised learning which exploits short-term memory. Meta-learning, often referred to as {\em 'learning to learn'} \cite{lemke2015metalearning}, focuses on optimizing the hyperparamters, i.e., parameters which dictate the learning algorithm, rather than being learned by it. In conventional non-meta-learned systems, these hyperparamters, also referred to as inductive bias, are handcrafted by human designers, typically based on abstract guidelines and experience. Meta-learning enables the learning of these parameters  without relying on heuristics.

\begin{figure}
    \centering
    \includegraphics[width=0.9\columnwidth]{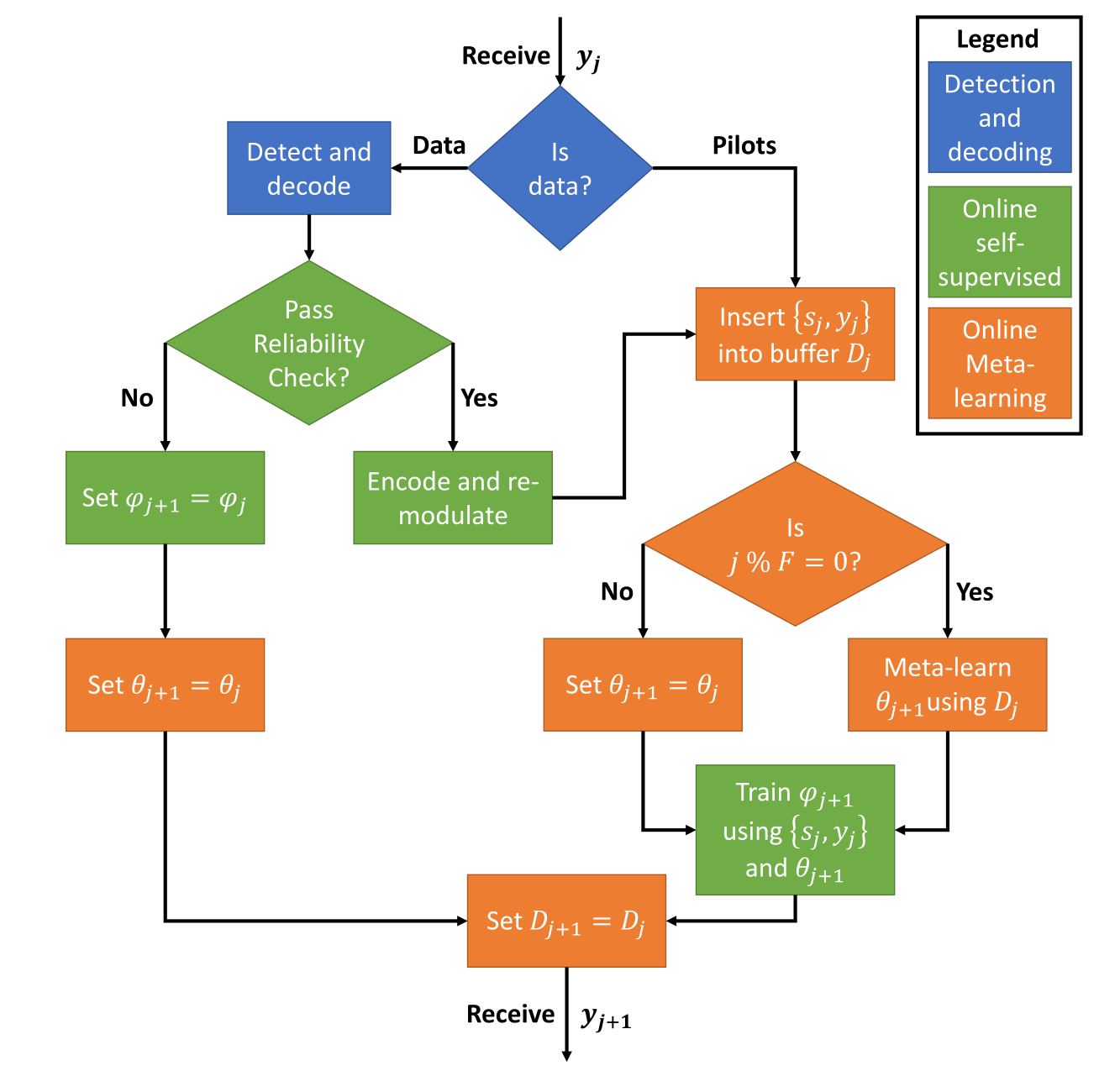}
    \caption{Illustration of the operation of the proposed meta-training algorithm in test phase.}
    \label{fig:FlowChart1}
\end{figure}

%This allows us to account for the predictive nature of self-supervised online training, where data from the current block is used to train for the subsequent (predicted) channel realization. Incorporating patterns in the channel variations, observed in past blocks, into the predictive detection aids to minimize the overall \ac{ber}.

% \textcolor{red}{Conclude this section by revealing the rationale of our approach to the readers (before digging into the details. This includes two important aspects of our algorithm: \\
% First, that instead of directly applying conventional \ac{maml}, as done in \cite{park2020learning}, we modify its operation to account for the predictive nature of self-supervised online training, where data from the current block is used to train for the subsequent (predicted) channel realization.   \\
% Second, that we use long-term variations in the channel to determine an initialization for the training algorithms which is expected to facilitate rapid training based on channel variation patters between subsequent blocks observed in the past.} 

To this end, following the  \ac{maml} framework \cite{finn2017model}, we define the support task as the {\em detection of the current block}; while the query task, for which the parameters should adapt, is the {\em detection of the next block}. 
Upon the reception of a block of channel outputs $\myVec{y}_j$, the method operates in three stages, illustrated as a block diagram in Fig.~\ref{fig:FlowChart1}:
\begin{enumerate}
    \item \emph{Detection}: Each incoming data block $\myVec{y}_j$ is mapped into an estimated symbol block $\hat{\myVec{s}}_j$, with the mapping determined by the current parameters vector $\Weights_{j}$. Then, it is decoded by using an \ac{fec} decoder to produce the demodulated and decoded message $\hat{\myVec{u}}_j$. When decoding is correct, as determined by error detection, the message $\hat{\myVec{u}}_j$ is re-encoded and modulated, producing an estimated transmitted vector $\tilde{\myVec{s}}_j$ (see Section~\ref{subsec:Online}). This block is inserted along with its observations $\myVec{y}_j$ into a labelled buffer $\mathcal{D}_{j}$.  This buffer contains pairs of previously received blocks $\myVec{y}_j$ along with their corresponding transmitted signal $\myVec{s}_j$, or an estimated version thereof. The buffer $\mathcal{D}_j$ contains $D$ such pairs, and is managed in a first-in-first-out mode. A pilot block $(\myVec{s}_j, \myVec{y}_j)$ is directly inserted into  $\mathcal{D}_{j}$ upon reception.
    \item \emph{Online training}: In each data block $j$, if decoding is successful, the weights $\Weights_{j+1}$ are updated by using the hyperparameters $\HyperParams_{j+1}$ and the newly decoded block $(\myVec{s}_{j}, \myVec{y}_{j})$, as detailed in Subsection~\ref{subsec:Online}. Otherwise, no update is carried out. A similar update takes place for pilot block $j$ with pilot block $(\myVec{s}_j, \myVec{y}_j)$. This update is the same as in \cite{shlezinger2019viterbinet}.
    \item \emph{Meta-learning}: The hyperparameter {$\HyperParams_{j+1}$} is optimized  to enable fast and efficient adaptation of the model parameter $\Weights_{j+1}$ based on the last successfully decoded block $(\myVec{s}_{j}, {\myVec{y}}_{j})$ using \eqref{eq:local_update}. Adopting MAML \cite{finn2017model}, we leverage the data in the buffer $\mathcal{D}_j$ by considering the problem
    %\NirCmt{I think that the indexing in \eqref{eq:meta_2} is confusing and inconsistent. I think it should be $ \HyperParams_{j +1}  =  \mathop{\arg \min}\limits_{\HyperParams}     \sum\limits_{\{(\myVec{s}_{j'}, {\myVec{y}}_{j'}), (\myVec{s}_{j'+1}, {\myVec{y}}_{j' +1})\} \in \mathcal{D}_j}     \mySet{L}_{j' +1}(\Weights =  \HyperParams  -  \kappa \nabla_{\HyperParams}\mySet{L}_{j'}(\HyperParams))$. Can you please double check?}:
    %
\begin{align}
    \label{eq:meta_2}
      \HyperParams_{j +1}  =  \mathop{\arg \min}\limits_{\HyperParams}     \sum\limits_{\{(\myVec{s}_{j'}, {\myVec{y}}_{j'}), (\myVec{s}_{j'+1}, {\myVec{y}}_{j' +1})\} \in \mathcal{D}_j}     \mySet{L}_{j' +1}(\Weights =  \HyperParams  -  \kappa \nabla_{\HyperParams }\mySet{L}_{j'}(\HyperParams)),
\end{align}
% %
% \begin{align}
%     \label{eq:meta_2}
%       \HyperParams_{j +1}  =  \mathop{\arg \min}_{\HyperParams}     \sum_{\{\myVec{s}_{j+1}, {\myVec{y}}_{j +1}\} \in \mathcal{D}_j}     \mySet{L}_{j +1}(\Weights_{j+1} =  \HyperParams  -  \kappa \nabla_{\HyperParams}\mySet{L}_{j}(\HyperParams)),
% \end{align}
where $\kappa >0$ is the meta-learning rate. The parameters $\Weights$ in \eqref{eq:meta_2} follow the same update rule in \eqref{eq:local_update} by using the last available block $(\myVec{s}_{j'}, \myVec{y}_{j'})$ in the buffer  prior to index $j'+1$. Furthermore, in line with \eqref{eq:meta_1}, the loss $\mySet{L}_{j'+1}(\Weights)$ is computed based on data from the {\em following} available block $(\myVec{s}_{j'+1}, \myVec{y}_{j'+1})$. When the buffer $\mathcal{D}_j$ contains a sufficiently diverse set of pairs of subsequent past channel realizations, the hyperparameter obtained via \eqref{eq:meta_2} should facilitate fast training for future channels via \eqref{eq:local_update} \cite{park2020end}. 
\end{enumerate} 

Online-meta training is applied  once every $F$ blocks with $I_{\rm meta}$ iterations. Moreover, if $\Weights_j$ and/or $\HyperParams_j$ are not updated in a given block index $j$, they are preserved for the next block by setting  $\Weights_{j+1} = \Weights_j$ and/or $\HyperParams_{j+1}=\HyperParams_j$.
The online adaptation framework %is detailed in the sequel, and 
is summarized in Algorithm~\ref{alg:online-meta-learning}.

\begin{algorithm} %!
\DontPrintSemicolon
\caption{Online Adaptation on Incoming Block $j$}
\label{alg:online-meta-learning}
\KwIn{Step sizes $\eta, \kappa$; number of meta-iterations $I_{\rm meta}$; threshold $\epsilon$; buffer $\mathcal{D}_{j}$; hyperparameter $\HyperParams_{j}$}
\KwOut{Hyperparameter $\HyperParams_{j+1}$; weights $\Weights_{j+1}$; buffer $\mathcal{D}_{j+1}$}
\vspace{0.15cm}
\hrule
\vspace{0.15cm}
%{\bf Initialization} inductive bias $\HyperParams_1$ and ViterbiNet $\Weights_1$ \\
%{\bf Initialization} buffer $\mathcal{D}_1$ as empty, i.e., $\mathcal{D}_1 \leftarrow [ \hspace{0.2cm} ]$ \\
%\For{$j = 1,2,\ldots$}{
Receive ${\myVec{y}}_j$ 	\tcp*{received channel output}
\uIf{Pilot}{
$\mathcal{D}_{j} \leftarrow \mathcal{D}_j \bigcup\{(\myVec{s}_j, {\myVec{y}}_j)\}$ \tcp*{known pilots} 
  }
  \uElse{
  Equalize and decode ${\myVec{y}}_j$ into  $\hat{\myVec{u}}_j$ \tcp*{data}
  \uIf{Decoding is correct}{
  Modulate  $\hat{\myVec{u}}_j\mapsto\myVec{s}_j$ \; 
$\mathcal{D}_{j} \leftarrow \mathcal{D}_j \bigcup\{(\myVec{s}_j, {\myVec{y}}_j)\}$ 
  }
%   \uElse{pass
%   }
  }
\nonl\texttt{Online predictive meta-learning (every $F$ blocks)}\;
Set $\HyperParams_{j+1}^{(0)} = \HyperParams_j$\; \label{stp:onlineStart}
\For{$i=0,1,\ldots$}{
\label{stp:Random} Randomly select block $(\myVec{s}_{j'+1}, {\myVec{y}}_{j'+1})\in \mathcal{D}_{j}$\; % except for the last block in the buffer $\mathcal{D}_j$\\  
      {\uIf{$(\myVec{s}_{j'}, {\myVec{y}}_{j'})\notin \mathcal{D}_{j}$}{
% pass;  
%   }
%   \uElse{
  go back to line \ref{stp:Random} \tcp*{invalid data for meta-learning}
  }
  }
  Set $\hat{\HyperParams}^{(0)} = \HyperParams_{j+1}^{(i)}$\;
    \For{$t=0,1,\ldots,I_{\rm meta}-1$}{
   Use block $(\myVec{s}_{j'}, {\myVec{y}}_{j'})$ to compute \tcp*{support task}
      \begin{equation*}
\hat{\Weights} = \hat{\HyperParams}^{(t)} -\eta\nabla_{\hat{\HyperParams}^{(t)}} \mySet{L}_{j'}(\Weights=\hat{\HyperParams}^{(t)}). \end{equation*} \;
    Use subsequent block $(\myVec{s}_{j'+1}, {\myVec{y}}_{j'+1})$ to update \tcp*{query task}
  \begin{equation*}
\hat{\HyperParams}^{(t+1)} = \hat{\HyperParams}^{(t)} - \kappa \nabla_{\hat{\HyperParams}^{(t)}}\mySet{L}_{j'+1}(\Weights =\hat{\Weights}). 
\end{equation*}
    }
    Update hyperparameter as $\HyperParams_{j+1}^{(i+1)} =\hat{\HyperParams}^{(I)}$ \tcp*{meta-update}
% Update the parameters with $I$ iterations on the selected block $(\myVec{s}_{j'}, {\myVec{y}}_{j'})$ via \eqref{eq:local_update} as %\tcp*{local update}
%   \begin{equation*}
% \hat{\Weights}_{j+1} = \HyperParams_{j+1}^{(i)}-\eta\nabla_{\HyperParams_{j+1}^{(i)}} \mySet{L}_{j}(\HyperParams_{j+1}^{(i)}). \end{equation*} \;
% Evaluate loss at block $j+1$% $\hat{\Weights}_{j+1}$ for block $j+1$ computed as
% , $\mySet{L}_{j+1}(\hat{\Weights}_{j+1})$\;
% Update hyperparameter $\HyperParams_{j+1}$ via \tcp*{meta-update}
%   \begin{equation*}
% \HyperParams_{j+1}^{(i+1)} = \HyperParams_{j+1}^{(i)} - \kappa \nabla_{\HyperParams_{j+1}^{(i)}}\mySet{L}_{j+1}(\hat{\Weights}_{j+1}). 
% \end{equation*}
}
Set hyperparameter $\HyperParams_{j+1} =\HyperParams_{j+1}^{(i+1)}$ \;
\nonl\texttt{Online learning (on each block)} \;
{\uIf{(Pilot) or (Decoding is correct)}{

 Train $\Weights_{j+1}$ with $({\myVec{s}}_j, {\myVec{y}}_j)$ and initialization $\HyperParams_{j+1}$  via \eqref{eq:meta_1} \tcp*{update} 
  }
  \uElse{ 
      $\Weights_{j+1} \leftarrow \Weights_j$ \tcp*{no update}
  }
  $\mathcal{D}_{j+1}\leftarrow \mathcal{D}_{j}$ \label{stp:onlineEnd} \tcp*{keep buffer}
  }
 %}
\end{algorithm}

% TODO NIR CONTINUE FROM HERE 13-Dec
%-----------------------------------
%	Discussion
%-----------------------------------
\vspace{-0.2cm}
\subsection{Discussion} 
\label{subsec:MetaLearningDiscussion}
\vspace{-0.1cm}
Algorithm~\ref{alg:online-meta-learning} is designed to enable rapid online training from scarce pilots by simultaneously accounting for short-term variations, via online training, and long-term variations, using the proposed predictive meta-learning. The proposed method does not require the transmission of additional pilots during the test phase. Rather,  data is generated in a self-supervised way, employing the current \ac{dnn}-based receiver and \ac{fec} codes, similar to the online training scheme in \cite{shlezinger2019viterbinet}. This self-training mechanism adapts the \ac{dnn}-based receiver as the channel changes during test phase. 
%The method is designed for scenarios where the channel changes continuously; If the channel remains constant or changes only slightly, one rather adapt $\Weights$ only during training phase. 
While Algorithm~\ref{alg:online-meta-learning} is designed for tracking time-varying channels, it also maintains valid weights for static channels, being trained with data corresponding to the underlying channel.

Contrary to common meta-learning schemes, such as the one in \cite{park2020meta}, our method treats support and query batches from subsequent channels realizations rather than from the same channel realization. This approach
%, which is particularly designed for online training in time-varying channels, induces predictive abilities into the weights, as the detector needs to detect successfully a block that will be drawn from the next channel rather than the current one. This technique 
improves upon the online training method of \cite{shlezinger2019viterbinet}, which assumes the subsequent channel to bear similarity to the current one, without accounting for temporal variation patterns observed in the past.

The gains associated with  Algorithm~\ref{alg:online-meta-learning} come at the cost of additional per-block complexity, which can be controlled by modifying the number of meta-learning iterations $I_{\rm meta}$,  and/or by changing its frequency, $F$. Furthermore, the rapid growth and expected proliferation of dedicated hardware accelerators for deep learning applications \cite{chen2019deep} indicates that the  number of hardware-capable digital communication devices will increase. Such devices are expected to be able to carry out the needed computations associated with online meta-learning in real-time.

We note that in Algorithm~\ref{alg:online-meta-learning} the number of iterations is given by constants $I_{\rm sgd}$ and $I_{\rm meta}$ for online and meta-learning, respectively. Accordingly, these values should be set so to allow convergence for both slow-varying or rapidly-varying channel settings (see Section~\ref{sec:Simulation}). More generally, one could also choose these parameters adaptively using as many iterations as needed. Depending on the current channel conditions, a possible way to realize this type of approach is to employ an early stopping policy, which terminates the training process when the loss decreases only slightly from one update to the next.	
	% ------------------------
	%	Model-Based Receivers
	%----------------------------------------------------------------------------------------
	%\vspace{-0.2cm}
	\section{Modular Training}
\label{sec:modular-training}

The training algorithm detailed in Section~\ref{sec:Hybrid Meta-Learning for Data-Driven Receivers} aids \ac{dnn}-based receivers to adapt in time-varying channels, reducing the overall transmission error rate. Nonetheless, the aforementioned training method may still struggle when applied to adapting highly-parameterized deep receivers to rapidly time-varying channels.  To facilitate adaptation and further minimize the overall error rate over the varying channels, we focus on deep receivers that utilize hybrid model-based/data-driven architectures \cite{farsad2021data, balatsoukas2019deep, shlezinger2020model}. We specifically modify the meta-learning method in Section~\ref{sec:Hybrid Meta-Learning for Data-Driven Receivers} to allow for different levels of adaptation per each module. 
%After proposing general modular training rule, by noting that model-based algorithms often have particular loss function for each module, we propose 
%This allows the method from Section~\ref{sec:Hybrid Meta-Learning for Data-Driven Receivers} to be applied individually to each such module. 
We formulate this approach mathematically in 
Subsection~\ref{subsec:modular_training_formulation}, with Subsection~\ref{subsec:modular_training_example} providing a specific instantiation of the proposed methodology based on the DeepSIC architecture \cite{shlezinger2019deepSIC}. We end with a discussion in Subsection~\ref{subsec:modular_training_discussion}.

\vspace{-0.2cm}
\subsection{Modular Training}
\label{subsec:modular_training_formulation}
\vspace{-0.1cm}
A modular model-based deep receiver is partitioned into separate modules, with each module $m=1,\ldots,M$ being specified by a parameter vector $\Weights^m$. Each module generally carries out a specific functionality within the communication receiver. We note that some of these functionalities require rapid adaptation, while other can be kept unchanged over a longer time scale. Our goal is to leverage this modular structure by means of meta-learning. We refer to the set of dynamic modules requiring adaptation as $\mathcal{M}_D$ and the set of stationary modules is denoted as $\mathcal{M}_S$. To this end, we modify problem \eqref{eq:meta_2} as
% and hyperparameter vector $\HyperParams$ can be divided into meaningful modules with parameter vectors $\Weights^m$ and hyperparameter vectors $\HyperParams^m$ for each module $m=1,\ldots,M$ (e.g., $\Weights = [\Weights^1, \ldots, \Weights^M]^\top$ and $\HyperParams = [\HyperParams^1, \ldots,\HyperParams^M]^\top$). Every module captures a specific aspect of the communication scenario. We group the modules that encounter rapid channel variations and collect the corresponding indices to form a set $\pi$; while indices of modules that exhibit static channels are denoted by the set $\pi^c$. Our goal is to employ this \emph{a priori} known information $\pi$ to find parameters and hyperparameters that enable rapid adaptation for time-varying channels; while parameters for $\pi^c$ modules remain unchanged. This can be formulated via (cf. \eqref{eq:meta_2})
\begin{multline}
    \label{eq:meta2_modular_general}
      \HyperParams_{j +1}  =  \mathop{\arg \min}\limits_{\HyperParams}     \sum\limits_{\{(\myVec{s}_{j'}, {\myVec{y}}_{j'}), (\myVec{s}_{j'+1}, {\myVec{y}}_{j' +1})\} \in \mathcal{D}_j}     \mySet{L}_{j' +1}([\Weights^1 =  \HyperParams^1 -  \kappa^1 \nabla_{(\HyperParams^1) }\mySet{L}_{j'}(\HyperParams), \ldots,\\,\ldots, \Weights^M =  \HyperParams^M -  \kappa^M \nabla_{(\HyperParams^M) }\mySet{L}_{j'}(\HyperParams)]^\top), 
\end{multline}
where we set a learning rate $\kappa^m > 0$ for $m \in \mathcal{M}_D$ and $\kappa^m = 0$ for $m \in \mathcal{M}_S$. This approach is similar to \cite{raghu2019rapid}, which considers a non-zero learning rate only for the last layer of a neural network, while positing that earlier layers do not need task-specific adaptations.

Since problem \eqref{eq:meta2_modular_general} has a high computational complexity when the number of modules is large, we propose to optimize each module separately by defining proper \emph{module-wise proximal loss} $\hat{\mySet{L}}$. This loss function is often available in hybrid model-based/data-driven approaches since generally, the distinct functional role of the modules is known in advance.

% {Precisely, we modify \eqref{eq:meta2_modular_general} by
%and propose to meta-learn only those parameters $\Weights^m$ and hyperparameters vectors $\HyperParams^m$  whose indice is in $\pi$. For model-based architectures, we suggest to replace \eqref{eq:local_update} by
% \begin{align}
    % \label{eq:meta2_modular_general_static}
    %   \Weights_{j +1}^m  =  \mathop{\arg \min}\limits_{\Weights^m}     \sum\limits_{\{(\myVec{s}_{j'}, {\myVec{y}}_{j'})\} \in \mathcal{D}_j}     \hat{\mySet{L}}^m_{j' +1}(\Weights^m),
% \end{align}
% for modules under static channels $m \in \pi^c$;

Specifically, in training phase, we initialize the overall parameter vector $\Weights = [\Weights^1, \ldots, \Weights^M]^\top$ for all $M$ modules by training over the obtained $T_p$ pilot blocks via \eqref{eq:meta_1} and \eqref{eq:local_update}. Then, online meta-learning takes place only for the dynamic modules in the set $\mathcal{M}_D$ by addressing the problem
%during the evaluation phase, we adopt the dynamic modules in set $\mathcal{M}_D$ by addressing the problem
\begin{align}
    \label{eq:meta2_modular_general_varying}
      \HyperParams_{j +1}^m  =  \mathop{\arg \min}\limits_{\HyperParams^m}     \sum\limits_{\{(\myVec{s}_{j'}, {\myVec{y}}_{j'}), (\myVec{s}_{j'+1}, {\myVec{y}}_{j' +1})\} \in \mathcal{D}_j}    \hat{\mySet{L}}_{j' +1}(\Weights_{j'+1}^m = \HyperParams^m - \kappa^m \nabla_{\HyperParams^m}  \hat{\mySet{L}}_{j'}(\HyperParams^m)),
\end{align}
while for the static modules in the set $\mathcal{M}_S$, we set
% \begin{align}
% \label{eq:meta2_modular_general_static}
     $ \HyperParams_{j +1}^m  =  \Weights_j^m$.
%\end{align}
Thus, during the evaluation phase, we only adapt the dynamic modules while the static module are obtained  from the training phase i.e., $\Weights_j^m = \Weights_{T_p}^m$ for $m \in \mathcal{M}_S$. The overall scheme is summarized in Algorithm~\ref{alg:modular-training}.

% Note that once \eqref{eq:meta2_modular_general_static} and \eqref{eq:meta2_modular_general_varying} are sufficiently learned, we can fix $\Weights^m$ for $m \in \pi^c$ and $\HyperParams^m$ for $m \in \pi$ \SangwooQuery{[is it true that online Meta-DeepSIC do not retrain $\Weights$ for static modules?]} while $\Weights^m$ for $m \in \pi$ should be adapted periodically to account for rapidly varying channels.
In the following, we exemplify the architecture described in this section for the problem of \ac{mimo} detection in uplink multi-user systems via soft interference cancellation (\ac{sic}).

\iffalse
\begin{align}
\label{eq:local_update_modular}
(\Weights^m)_{j+1}^{(t+1)} = (\Weights^m)_{j+1}^{(t)} - \eta \nabla_{\Weights^m}\mySet{L}_{j}((\Weights^m)_{j+1}^{(t)}), \: m \in \pi
 \end{align}
 \eqref{eq:meta2_modular_general} as
\begin{align}
    \label{eq:meta2_modular}
      \HyperParams^m_{j +1}  =  \mathop{\arg \min}\limits_{\HyperParams^m}     \sum\limits_{\{(\myVec{s}_{j'}, {\myVec{y}}_{j'}), (\myVec{s}_{j'+1}, {\myVec{y}}_{j' +1})\} \in \mathcal{D}_j}     \mySet{L}_{j' +1}(\Weights^m =  \HyperParams^m  -  \kappa \nabla_{(\Weights^m) }\mySet{L}_{j'}(\HyperParams^m)),  \: m \in \pi.
\end{align}
\fi

\begin{algorithm} %!
\DontPrintSemicolon
\caption{Modular Adaptation on Incoming Block $j$ (Steps \ref{stp:onlineStart}-\ref{stp:onlineEnd} of Algorithm~\ref{alg:online-meta-learning})}
\label{alg:modular-training}
\KwIn{Step sizes $\eta, \kappa$; number of meta-iterations $I_{\rm meta}$; threshold $\epsilon$; buffer $\mathcal{D}_{j}$; hyperparameter $\HyperParams_{j}$; rapidly-changing indices $\pi$; static modules indices $\pi^c$}
\KwOut{Hyperparameter $\HyperParams_{j+1}$; weights $\Weights_{j+1}$; buffer $\mathcal{D}_{j+1}$}
\vspace{0.15cm}
\hrule
\vspace{0.15cm}
%{\bf Initialization} inductive bias $\HyperParams_1$ and ViterbiNet $\Weights_1$ \\
%{\bf Initialization} buffer $\mathcal{D}_1$ as empty, i.e., $\mathcal{D}_1 \leftarrow [ \hspace{0.2cm} ]$ \\
%\For{$j = 1,2,\ldots$}{
\nonl\texttt{Online predictive meta-learning (every $F$ blocks)}\;
\For{$m = 1,\ldots,M$}{
\uIf{$m \in \pi$ }{ 
Set $\HyperParams_{j+1}^{(0)} = \HyperParams^m_j$\;
\For{$i=0,1,\ldots$}{
\label{stp:Random2} Randomly select block $(\myVec{s}_{j'+1}, {\myVec{y}}_{j'+1})\in \mathcal{D}_{j}$\; % except for the last block in the buffer $\mathcal{D}_j$\\  
      {\uIf{$(\myVec{s}_{j'}, {\myVec{y}}_{j'})\notin \mathcal{D}_{j}$}{
  go back to line \ref{stp:Random2} \tcp*{invalid data for meta-learning}
  }
  }
  Set $\hat{\HyperParams}^{(0)} = \HyperParams_{j+1}^{(i)}$\;
    \For{$t=0,1,\ldots,I_{\rm meta}-1$}{
   Use block $(\myVec{s}_{j'}, {\myVec{y}}_{j'})$ to compute \tcp*{support task}
      \begin{equation*}
\hat{\Weights} = \hat{\HyperParams}^{(t)} -\eta\nabla_{\hat{\HyperParams}^{(t)}} \hat{\mySet{L}}_{j'}(\Weights^m=\hat{\HyperParams}^{(t)}). \end{equation*} \;
    Use subsequent block $(\myVec{s}_{j'+1}, {\myVec{y}}_{j'+1})$ to update \tcp*{query task}
  \begin{equation*}
\hat{\HyperParams}^{(t+1)} = \hat{\HyperParams}^{(t)} - \kappa^m \nabla_{\hat{\HyperParams}^{(t)} }\hat{\mySet{L}}_{j'+1}(\Weights^m =\hat{\Weights}). 
\end{equation*}
    }
    Update hyperparameter as $\HyperParams_{j+1}^{(i+1)} =\hat{\HyperParams}^{(I)}$ \tcp*{meta-update}
}
Set hyperparameter $\HyperParams^m_{j+1} =\HyperParams_{j+1}^{(i+1)}$ \tcp*{update the hyperparameter}
}
\uElse{ 
      Set hyperparameter $\HyperParams^m_{j+1} =\Weights^m_{j}$ \tcp*{copy the $m$th hyperparameter}
  }
}

\nonl\texttt{Online learning (on each block)} \;
{\uIf{(Pilot) or (Decoding is correct)}{
 \For{$m \in \pi$}{Train $\Weights^m_{j+1}$ with $({\myVec{s}}_j, {\myVec{y}}_j)$ and initialization $\HyperParams^m_{j+1}$  via \eqref{eq:meta_1} \tcp*{update} }
 \For{$m \in \pi^c$}{
 $\Weights^m_{j+1} \leftarrow \Weights^m_j$ \tcp*{no update for static} }
  }
  \uElse{ 
      $\Weights_{j+1} \leftarrow \Weights_j$ \tcp*{no update}
  }
  $\mathcal{D}_{j+1}\leftarrow \mathcal{D}_{j}$ \tcp*{keep buffer}
  }
 %}
\end{algorithm}

% \begin{align}
%       \HyperParams^*  =  \mathop{\arg \min}\limits_{\HyperParams = [\HyperParams^1, \ldots, \HyperParams^M]^\top}     \mathbb{E}_{P_\tau} \left[ \mySet{L}_{\tau}(\Weights =  [\HyperParams^1  -  \kappa^1 \nabla_{\HyperParams^1 }\mySet{L}_{\tau}(\HyperParams),\ldots, \HyperParams^M  -  \kappa^M \nabla_{\HyperParams^M }\mySet{L}_{\tau}(\HyperParams)]^\top)\right],
%       \label{eq:general_anil}
% \end{align}
% Our approach, coined modular training, seeks to jointly exploit these properties for  setups  where  the  interpretable  nature  of  the  deep  receiver  sub-blocks  can  be  related  with the  components  of  the  channel  where  variations  originate. Specifically, as hybrid model-based/data-driven receivers can often divided into meaningful sub-blocks, we wish to train only those that relate to a specific time-varying component of the underlying channel. 
% We next show how to apply modular training in uplink \ac{mimo} systems, employing the DeepSIC receiver \cite{shlezinger2019deepSIC}.

\vspace{-0.2cm}
\subsection{Modular DeepSIC Receiver}
\label{subsec:modular_training_example}
\vspace{-0.1cm}
We focus on a special case of the generic  model detailed in Subsection~\ref{subsec:Channel}, in which the channel is memoryless ($L=1$), while the receiver has $N>1$ antennas and it communicates with $K>1$ users. A single user with index $k'\in \{1,\ldots,K\}$ is mobile, while the remaining are static. We assume that the receiver knows which of the users is mobile and which is static, i.e., it has prior knowledge of $k'$. Such scenario arise, for instance, in a traffic monitoring infrastructure, where a multi-antenna \acl{bs} communicates with both static road-side units and mobile vehicles. Generalizations can be directly obtained  from the method described here.

\begin{figure}
	\centering
	\includegraphics[width = 0.9\columnwidth]{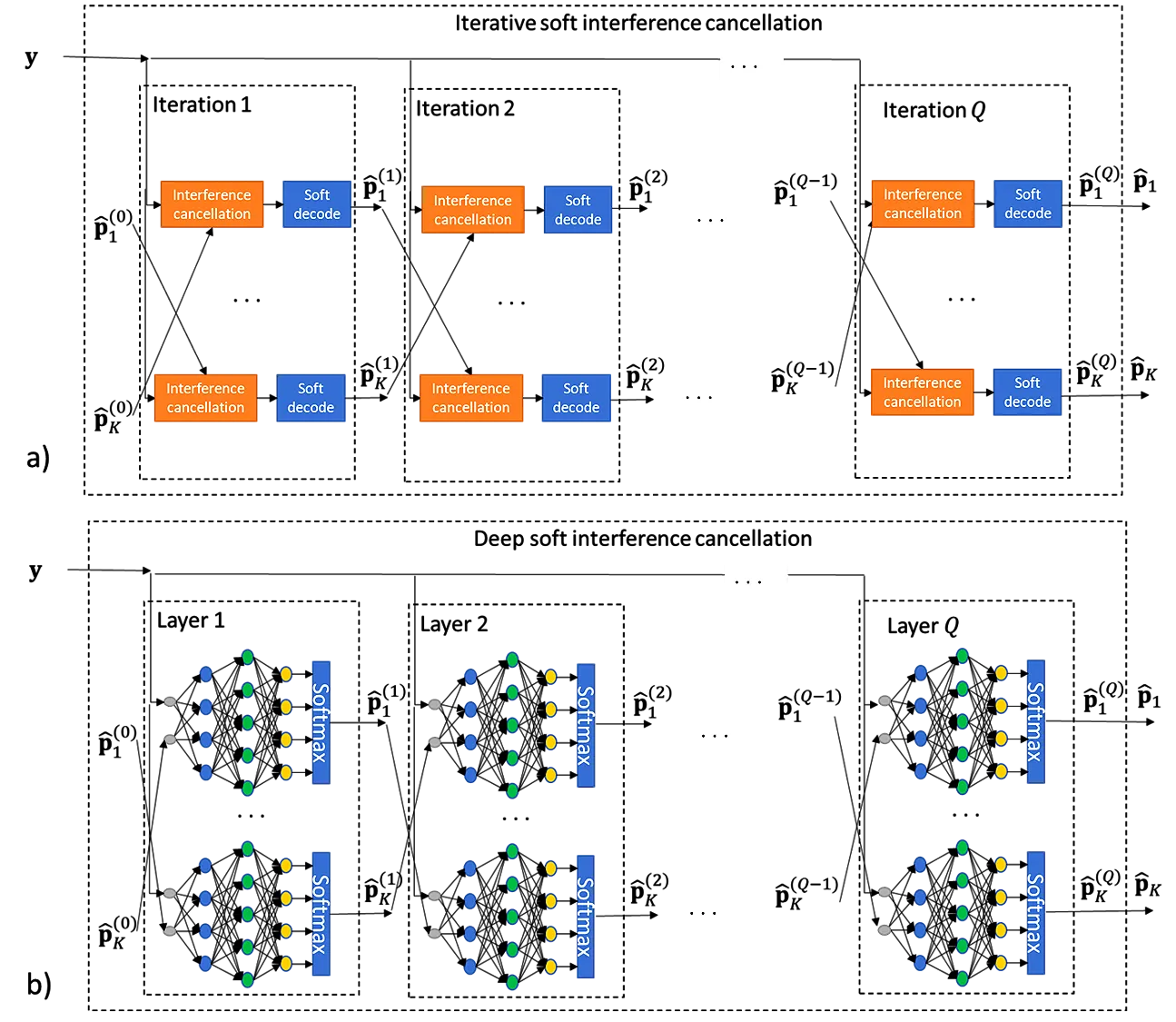}
	\vspace{-0.4cm}
	\caption{(a) Iterative soft interference cancellation, and (b) the \ac{dnn}-based DeepSIC.}
	%	\vspace{-0.4cm}
	\label{fig:SoftIC}
\end{figure}

We adopt the DeepSIC receiver proposed in \cite{shlezinger2019deepSIC} as a modular receiver architecture. DeepSIC is derived from iterative \ac{sic} \cite{choi2000iterative}, which is a \ac{mimo}  detection method combining multi-stage interference cancellation with soft decisions. For brevity, we omit here the subscripts representing the time instance and the block index, using $\myVec{Y}$ to denote the random channel output and $\myVec{S}=[\myVec{S}_1,\ldots,\myVec{S}_K]^T$ the transmitted symbols. 
DeepSIC operates in $Q$ iterations, refining an estimate of the conditional \acl{pmf} of $\myVec{S}_k$, denoted by $\hat{\myVec{p}}_{k}^{(q)}$ where $q$ is the iteration number. This estimate   is generated for every symbol $k \in \{1,2,\ldots, K\}$ by using the corresponding estimates of the interfering symbols $\{S_l\}_{l \neq k}$ obtained in the previous iteration $\{\hat{\myVec{p}}_{\tilde{k}}^{(q-1)}\}_{\tilde{k}\neq k}$. Iteratively repeating this procedure refines the conditional distribution estimates, allowing the detector to accurately recover each symbol from the output of the last iteration.  This iterative procedure is illustrated in Fig.~\ref{fig:SoftIC}(a). 

In DeepSIC, the interference cancellation and soft decoding steps are implemented with \acp{dnn}. The soft estimate of the symbol transmitted by $k$th user in the $q$th iteration is done by a \ac{dnn}-based classifying module with parameters $\Weights^{k,q}$. The output of the \ac{dnn} module with parameters $\Weights^{k,q}$ is a soft estimate of the symbol of the $k$th user. The set $\Weights = \{\Weights^{k,q}\}$ describes the model parameters. The resulting \ac{dnn}-based receiver, illustrated in Fig.~\ref{fig:SoftIC}(b), was shown in \cite{shlezinger2019deepSIC}  to accurately carry out \ac{sic}-based \ac{mimo} detection in complex channel models.

We employ modular training as described in Algorithm~\ref{alg:modular-training}. DeepSIC uses $M=K \cdot Q$ modules, with the $(k,q)$-th module detecting the symbol of user $k$ at iteration $q$. Since only user $k'$ is moving, we define the set of dynamic modules as $\mathcal{M}_D = \{(k',q)\:|\:q\in\{1,\ldots,Q\}\}$, with all other modules being static. We define the module-wise proximal loss \eqref{eq:meta2_modular_general_varying} for module $k$ at iteration $q$ as the cross entropy
\begin{align}
   \hat{\mySet{L}}_{j}(\Weights^{k,q}) = -\sum_{i=1}^{\Blklen}\log \hat{P}_{\Weights^{k,q}}(\myVec{y}_{i,j}, \{ \hat{\myVec{p}}_{i,j}^{\tilde{k},q-1} \}_{\tilde{k}\neq k} | \myStateR_{i,j}),
\end{align}
where $\hat{\myVec{p}}^{\tilde{k},q-1}$ is the estimated symbol probability for the $\tilde{k}$th user at iteration $q-1$. The module-wise parameter vector $\Weights_{j+1}^{k,q}$ for each $(k,q)\in \mathcal{M}_D$ is obtained from $I$ gradient descent iterations
\begin{align}
    \Weights_{j+1}^{k,q} = \HyperParams^{k,q}_{j+1} - \kappa^{k,q} \nabla_{ \HyperParams^{k,q}}\hat{\mySet{L}}_{j}(\Weights^{k,q}_{j+1})
    \label{eq:meta3_modular_general_varying_deepsic}
\end{align}
with $\HyperParams^{k,q}_{j+1}$ being optimized via \eqref{eq:meta2_modular_general_varying}, namely,
\begin{align}
    \label{eq:meta2_modular_general_varying_deepsic}
      \HyperParams_{j +1}^{k,q}  =  \mathop{\arg \min}\limits_{\HyperParams^{k,q}}     \sum\limits_{\{(\myVec{s}_{j'}, {\myVec{y}}_{j'}), (\myVec{s}_{j'+1}, {\myVec{y}}_{j' +1})\} \in \mathcal{D}_j}    \hat{\mySet{L}}_{j' +1}(\Weights^{k,q} = \HyperParams^{k,q} - \kappa^{k,q} \nabla_{ \HyperParams^{k,q}} \hat{\mySet{L}}_{j'}(\HyperParams^{k,q})).
\end{align}
We note that problem \eqref{eq:meta2_modular_general_varying_deepsic} uses all previous data up to time $j+1$; while \eqref{eq:meta3_modular_general_varying_deepsic} only uses only data from block $j$ for the local update.

\vspace{-0.2cm}
\subsection{Discussion}
\label{subsec:modular_training_discussion}
\vspace{-0.1cm}
Modular training via Algorithm~\ref{alg:modular-training} exploits the model-based structure of hybrid \ac{dnn} receivers, applying  Algorithm~\ref{alg:modular-training} to the subset of relevant modules in the overall  network, i.e., it uses the same data as in Algorithm~\ref{alg:online-meta-learning} while adapting less parameters, thus gaining in convergence speed and accuracy of the trained model. The exact portion of the overall parameters is  dictated  by  the  ratio  of  the mobile users to the overall users in the network. The smaller the ratio is, the more one can benefit from modular training. We validate the benefits of modular training in Subsection~\ref{subsec:modular-training} and show that this approach can decrease the average \ac{ber} significantly. Indeed, our experimental results demonstrate that the proposed approach successfully translates the prior knowledge on the nature of  the  variations  into  improved  performance,  particularly  when  dealing  with  short  blocks,  i.e. short coherence duration of rapidly varying channels. 

This modular training approach can be viewed as a form of transfer learning, as it trains online only parts of the parameters. Yet, it is specifically tailored towards interpretable architectures, adapting internal sub-modules of the overall network, as opposed to the common transfer learning approach of retraining the output layers. For instance, the suggested method trains the sub-modules associated with all mobile users separately in an iterative fashion, building upon the fact that each module $\Weights^{k,q}$ should produce a soft estimate of $\myVec{S}_k$.  Modular training can also be enhanced to carry out some final tuning of the overall network since, e.g., the channels of the mobile users is expected to also affect how their interference is canceled when recovering the remaining users via the sub-modules which are not adapted online.  
Finally, it is noted that this approach relies on knowledge of the mobile user, and thus incorrect knowledge of $k'$ is expected to affect the accuracy of the trained receiver.  We  leave  the  study  of  the resilience of modular training and the aforementioned extensions for future work.
	%----------------------------------------------------------------------------------------
	%	Numerical Evaluations
	%----------------------------------------------------------------------------------------
	%\vspace{-0.2cm}
	\section{Numerical Evaluations}
\label{sec:Simulation}
\vspace{-0.1cm} 
In this section we numerically evaluate the proposed online adaptation scheme in finite-memory \ac{siso} channels and in memoryless multi-user \ac{mimo} setups\footnote{The source code used in our experiments is available at \href{https://github.com/tomerraviv95/MetaDeepSIC}{https://github.com/tomerraviv95/MetaDeepSIC}}. 
We first describe the receivers compared in our experimental study, detailing the architectures in Subsection~\ref{subsec:simEqualizers} and the training methods in Subsection~\ref{subbsec:Training Methods}. Then, we present the main simulation results for evaluating online meta-learning on linear synthetic channels, non-linear synthetic channels, and channels obeying the COST 2100 model in Subsections~\ref{subsec:synth_simulation_results} to \ref{subsec:cost_simulation_results}. Finally, we numerically evaluate Algorithm~\ref{alg:modular-training} which combines online meta-learning with the modular training in Subsection~\ref{subsec:modular-training}.
%, as well as the effect of our learning schemes on pilot efficiency, in Subsections~\ref{subsec:modular-training} and \ref{subsec:pilot-training}, respectively.

%-----------------------------------
%	ViterbiNet
%-----------------------------------
% \subsection{ViterbiNet}
% \label{subsec:ViterbiNet}

% Following the channel definition of Subsection~\ref{subsec:SISOChannel}, the Viterbi equalizer solves the maximum likelihood sequence detection problem
% \begin{align}
% \hat{\myVec{s}}_j   
% &= \mathop{\arg \min}_{\myVec{s} \in \mySet{S} }\left\{-\sum\limits_{i=1}^{\Blklen } \log \Pdf{\myVec{Y}_{i,j} | \myState_{i,j}}\left( \myVec{y}_{i,j}  |  \myStateR_{i,j}\right)\right\}.
% \label{eqn:ML3} 
% \end{align}	
% In particular,  \eqref{eqn:ML3} is solved recursively via dynamic programming, by  iteratively updating a {\em path cost} $c_i(\myStateR)$ for each state $\myStateR\in \bar{\mySet{S}}$ for $i=1,2,\ldots,B$. ViterbiNet implements Viterbi detection in a data-driven fashion by training a \ac{dnn} to provide a parametric estimate of the likelihood function $\hat{P}_{\Weights}\left( \myVec{y}  |  \myStateR\right)$. Meta-ViterbiNet \cite{raviv2021meta} applies Algorithm~\ref{alg:online-meta-learning} to ViterbiNet, leading to higher adaptivity over fast channel variations.

%-----------------------------------
%	Evaluated Equalizers
%-----------------------------------
\vspace{-0.4cm}
\subsection{Evaluated Receivers}
\label{subsec:simEqualizers}
The receiver algorithms used in our experimental study are tailored to the considered settings:
%\vspace{-0.1cm}
\subsubsection{Finite-Memory \ac{siso} Channels}
We compare two  \ac{dnn}-based receivers for \ac{siso} settings:
\begin{itemize}
    \item The ViterbiNet equalizer, proposed in \cite{shlezinger2019viterbinet}, which is a \ac{dnn}-based Viterbi detector for finite-memory channels of the form \eqref{eqn:ChModel1} \cite{viterbi1967error}: The Viterbi equalizer (from which ViterbiNet originates) solves the maximum likelihood sequence detection problem
\begin{align}
\hat{\myVec{s}}_j   
&= \mathop{\arg \min}_{\myVec{s} \in \mySet{S} }\left\{-\sum\limits_{i=1}^{\Blklen } \log \Pdf{\myVec{Y}_{i,j} | \myState_{i-\Mem+1,j},\ldots, \myState_{i,j}}\left( \myVec{y}_{i,j}  |  \myStateR_{i-\Mem+1,j},\ldots, \myStateR_{i,j}\right)\right\}.
\label{eqn:ML3} 
\end{align}	
    ViterbiNet computes each log likelihood using a  \ac{dnn} that requires no  knowledge of the channel  distributions $ \Pdf{\myVec{Y}_j |  \myState_{i-\Mem+1,j},\ldots, \myState_{i,j}}$. This internal \ac{dnn} is implemented using three fully-connected layers of sizes $1\times 100$, $100\times 50$, and $50 \times |\mySet{S}|^\Mem$, with activation functions set to sigmoid (after first layer), ReLU (after second layer), and softmax output layer.
    %sigmoid and ReLU activations, respectively, and a softmax output layer. 
    \item A recurrent neural network  symbol detector, comprised of a sliding-window \ac{lstm} classifier with two hidden layers of 256 cells and window size $\Mem$, representing a black-box \ac{dnn} benchmark \cite{tandler2019recurrent}.
\end{itemize}
The Viterbi equalizer  with complete knowledge of the channel is used as a baseline method.

\subsubsection{Memoryless \ac{mimo} Channels}
We evaluate two  \ac{dnn}-based \ac{mimo}  receivers:
\begin{itemize}
    \item The DeepSIC receiver detailed in Subsection~\ref{subsec:modular_training_example}: The building blocks of DeepSIC are implemented using three fully-connected layers:  An $(N + K - 1) \times 100$ first layer, a $100 \times 50$ second layer, and a $50 \times |\mySet{S}|$ third layer, with a sigmoid and a ReLU intermediate activation functions, respectively. The number of iterations is set to $Q=5$.  
    %sigmoid and ReLU activations, respectively, and a softmax output layer. 
    \item A ResNet10 black-box neural network, which is based on the  DeepRX architecture proposed in \cite{honkala2021deeprx}: The architecture is comprised of 10 residual blocks \cite{he2016deep}, each block has two convolutional layers with 3x3 kernel, one pixel padding on both sides, and no bias terms, with a ReLU  in-between. 
    A 2D batch normalization follows each convolutional layer.
\end{itemize}

\begin{figure*}
    \centering
    \begin{subfigure}[b]{0.31\textwidth}
    \includegraphics[width=\textwidth,height=0.15\textheight]{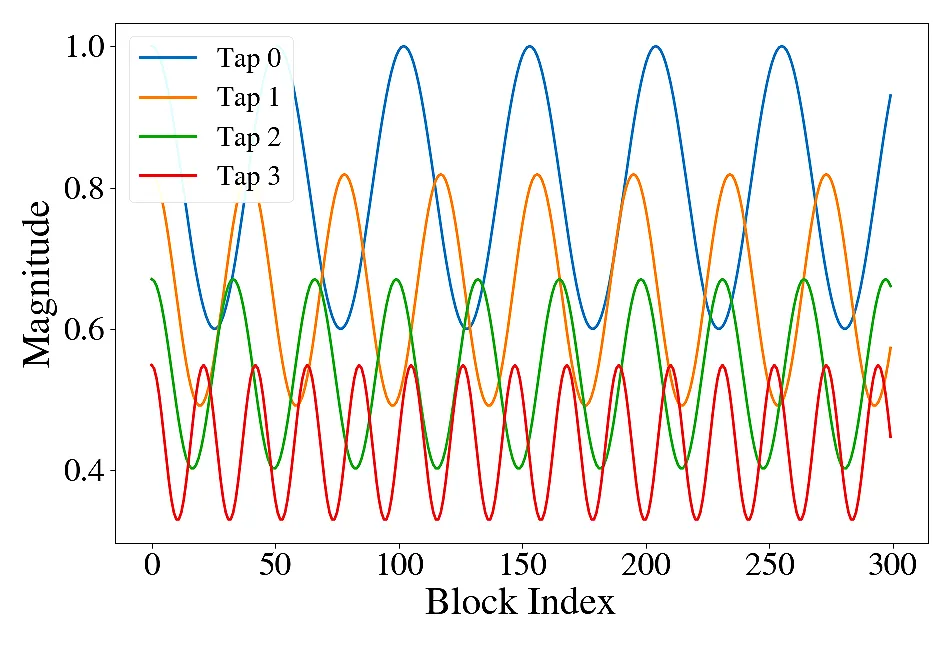}
    \caption{Synthetic training channel.}
    \label{fig:synthetic_train_channel}
    \end{subfigure} 
    \begin{subfigure}[b]{0.31\textwidth}
    \includegraphics[width=\textwidth,height=0.15\textheight]{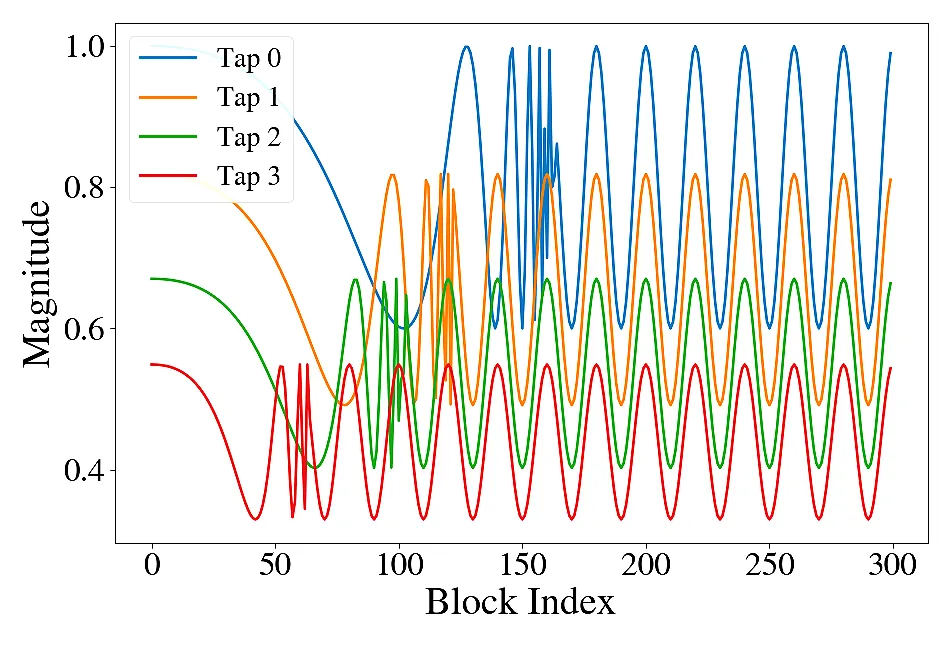}
    \caption{Synthetic test channel.}
    \label{fig:synthetic_test_channel}
    \end{subfigure}
    \begin{subfigure}[b]{0.31\textwidth}
    \includegraphics[width=\textwidth,height=0.15\textheight]{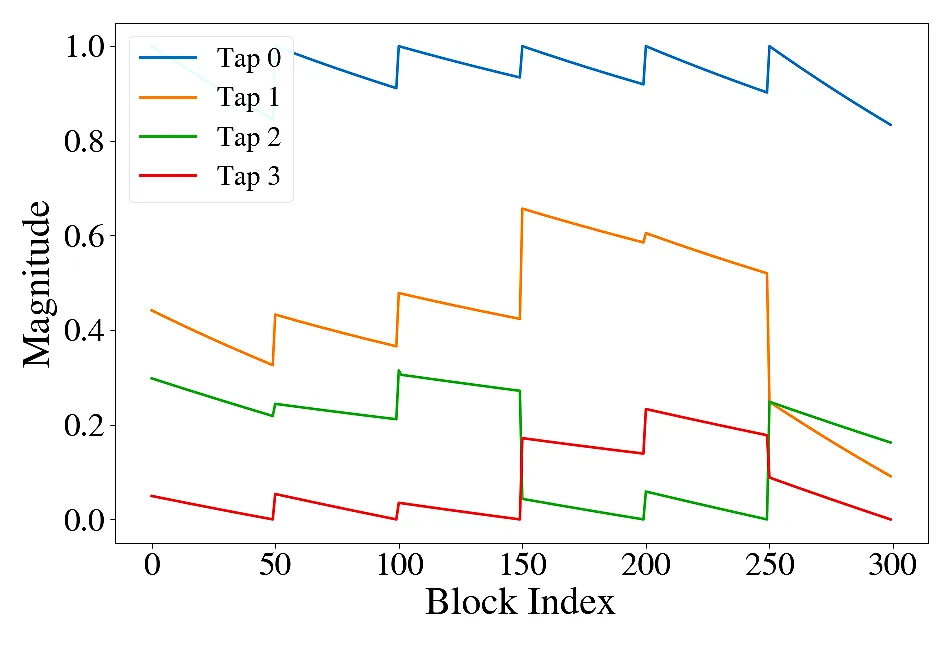}
    \caption{COST 2100 channel.}
    \label{fig:COSTChannel} 
    \end{subfigure}
    \caption{Examples of time-varying channels: channel coefficients versus block index.}
    \figSpace
    \label{fig:channels} 
 %   \vspace{-0.1cm}
\end{figure*}

\subsection{Training Methods}
\label{subbsec:Training Methods}
Before the test phase begins, we generate a set $\mathcal{D}_0$ of $T_p$ pilot blocks. We then use the following methods for adapting the deep receivers:
\begin{itemize}
    \item {\em Joint training}: The \ac{dnn} is  trained on $\mathcal{D}_0$ only by minimizing the empirical cross-entropy loss, and no additional training is performed in the test phase.
    \item {\em Online training} \cite{shlezinger2019viterbinet}: The \ac{dnn} is initially trained  on $\mathcal{D}_0$ by minimizing the empirical cross-entropy loss. Then, during test, the \ac{dnn} parameter vector $\Weights_j$ is re-trained on each successfully decoded data block and on each incoming pilot block as specified in \eqref{eq:local_update}, using $\Weights_{j}$ in lieu of $\HyperParams_{j+1}$. %, using a fixed non-optimized inductive bias $\HyperParams$.
    \item {\em Online meta-learning}: 
    Here, we first meta-train $\HyperParams_0$ with $\mathcal{D}_0$ similar to \eqref{eq:meta_2} as
    \begin{align*}
    %\label{eq:offline_meta}
     \HyperParams_{0} = \mathop{\arg \min}_{\HyperParams} \sum_{\{(\myVec{s}_{ j'}, {\myVec{y}}_{ j'}), (\myVec{s}_{ j'+1}, {\myVec{y}}_{ j'+1})\} \in \mathcal{D}_0}    \mySet{L}_{ j'+1}(\Weights = \HyperParams  -  \kappa \nabla_{ \HyperParams }\mySet{L}_{ j'}(\Weights =\HyperParams)).
\end{align*}
    
    %To this end, meta-training splits the buffer $\mathcal{D}_0$ into following-words pairs as $(\boldsymbol{y}^B_0,\boldsymbol{y}^B_1),\ldots,(\boldsymbol{y}^B_{T_t-2},\boldsymbol{y}^B_{T_t-1})$. The first word of each pair $\boldsymbol{y}^B_{2n}$ is used for the local update (\ref{eq:local_update}), while the second word $\boldsymbol{y}^B_{2n+1}$ is used for the global update calculation as in
	%\begin{align}
	%\label{eq:meta_2_pred_offline}
	 % \mathop{\arg \min}_{\HyperParams_0} \sum_{\text{odd }j} \mySet{L}_j(\Weights= \HyperParams_0 - \eta \nabla_{ \HyperParams_0}\mySet{L}_{j-1}( \HyperParams_0)).
	%\end{align}
	This  process yields the initial hyperparameters $\HyperParams_0$. 
	Then, during test,   Algorithm~\ref{alg:online-meta-learning} is used with online learning every block and online meta-learning every $F=5$ blocks. The number of online meta-learning updates   equals that of online training, thus inducing a relative small overhead due to its additional computations. The combination of online meta-learning with ViterbiNet is henceforth referred to as {\em Meta-ViterbiNet}, while the DeepSIC receiver trained with this approach is coined {\em Meta-DeepSIC}.
\end{itemize}

All training methods use the Adam optimizer \cite{kingma2014adam} with $I_{\rm sgd}=200$ iterations and learning rate $\eta = 10^{-3}$; for meta-learning, we set the meta-learning rate to $\kappa = 10^{-1}$ and the iterations to $I_{\rm meta}=200$. Both meta-learning and online learning employ a batch size of $64$ symbols. These values were set empirically such that the receivers' parameters approximately converge at each time step (for both meta-learning and online learning). In case of the online training or meta-training schemes, re-training occurs if the normalized bits difference between the re-encoded word and the hard-decision of the channel word is smaller than the threshold of $0.02$. The complete simulation parameters can also be found in the source code available on GitHub.

%-----------------------------------
%	Simulation Results
%-----------------------------------
\vspace{-0.4cm}
\subsection{Linear Synthetic Channel Results}
\label{subsec:synth_simulation_results}
\vspace{-0.1cm}
We begin by evaluating Algorithm~\ref{alg:online-meta-learning} on synthetic time-varying linear channels with additive Gaussian noise, considering  a finite-memory \ac{siso} setting and a memoryless \ac{mimo} setup.
\subsubsection{\ac{siso} Finite-Memory Channels} 
%We refer to the combination of the ViterbiNet equalizer and online meta-training as Meta-ViterbiNet. 
Recalling Fig.~\ref{fig:transmission}, we transmit $T_p=T_d=300$ blocks comprised of $\Blklen = 136$ symbols, i.e., a relatively short coherence duration for the time-varying channel. Each block includes $120$ information bits, encoded using a \acl{rs} [17,15] code with two parity symbols with \ac{bpsk} modulation, i.e., $\mySet{S} = \{\pm 1\}$.  %The online meta-learning in Algorithm \ref{alg:online-meta-learning} is executed every $K=5$ blocks. 

We consider a linear Gaussian channel, whose input-output relationship is given by
\begin{equation}
\label{eqn:Gaussian}
\myY_{i,j} = \sum_{l=0}^{\Mem-1} h_{l,j}S_{i-l,j} + w_{i,j},
\end{equation}
where $\myVec{h}_j = [h_{0,j},\ldots,h_{\Mem-1,j}]^T$ are the real channel taps, and $w_{i,j}$ is \acl{awgn} with variance $\sigma^2$. The channel memory is $\Mem = 4$ with the taps $\{h_{l,j}\}$  generated using a synthetic model representing oscillations of varying frequencies. 
Here,  the signals received during the pilots used for initial training $(\mathcal{D}_0)$ are subject to the time-varying channel whose taps are illustrated in Fig.~\ref{fig:synthetic_train_channel}; while we use the taps  in Fig.~\ref{fig:synthetic_test_channel} for testing. This channel represents oscillations of varying frequencies. %, where  the periods of the taps become aligned as the noise subsides. 

\begin{figure*}
    \centering
    \begin{subfigure}[b]{0.48\textwidth}
    \includegraphics[width=\textwidth,height=0.22\textheight]{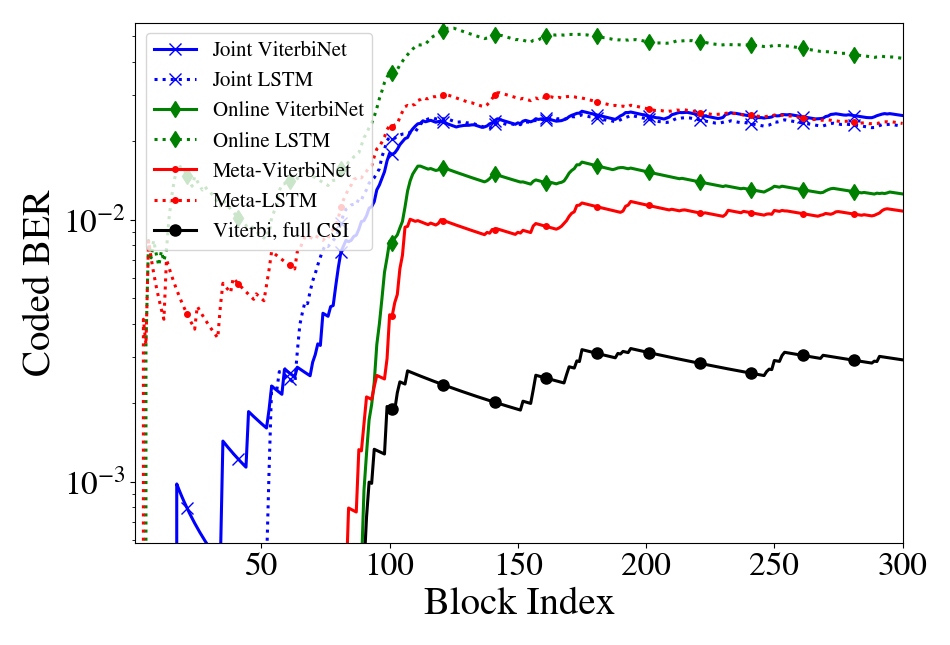}
    \caption{Coded \ac{ber} vs. block index, ${\rm SNR} = 12$ dB.}
    \label{fig:SyntheticBERvsBlock}
    \end{subfigure}
    \begin{subfigure}[b]{0.48\textwidth}
    \includegraphics[width=\textwidth,height=0.22\textheight]{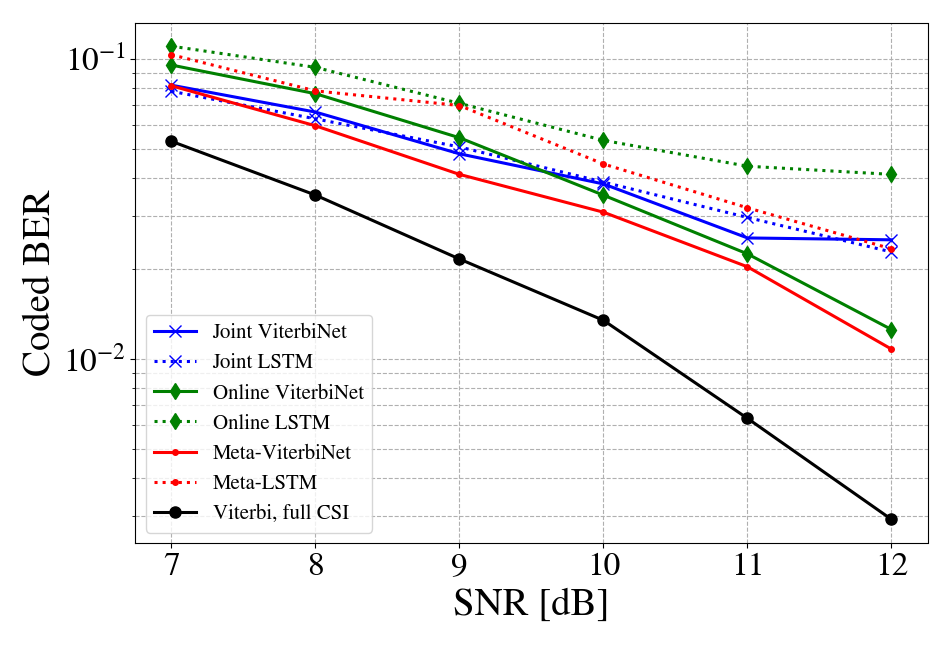}
    \caption{Coded \ac{ber} after 300 blocks vs. SNR.}
    \label{fig:SyntheticBERvsSNR}
    \end{subfigure}
    \caption{\ac{siso} synthetic linear Gaussian channel, $\Blklen = 136$.}
    \label{fig:SyntheticBER} 
    \figSpace
\end{figure*}

In Fig.~\ref{fig:SyntheticBERvsBlock} we plot the evolution of the average coded \ac{ber} of the considered receivers when the \ac{snr}, defined as $1/\sigma^2$, is set to $12$ dB. Fig.~\ref{fig:SyntheticBERvsBlock} shows that Meta-ViterbiNet significantly outperforms its benchmarks, demonstrating the gains of the proposed online meta-learning training and its suitability when combined with receiver architectures utilizing relatively compact \acp{dnn}. In particular, it is demonstrated that each of the ingredients combined in Meta-VitebiNet, i.e., its model-based architecture, the usage of self-supervision, and the incorporation of online meta-learning, facilitates operation in time-varying conditions: The ViterbiNet architecture consistently outperforms the black-box \ac{lstm} classifier; Online training yields reduced \ac{ber} as compared to joint learning; and its combination with meta-learning via Algorithm~\ref{alg:online-meta-learning} yields the lowest \ac{ber} values.  
To further validate that these gains also hold for different \acp{snr}, we show in Fig.~\ref{fig:SyntheticBERvsSNR} the average coded \ac{ber} of the evaluated receivers after 300 blocks, averaged over 5 trials. We observe in Fig.~\ref{fig:SyntheticBERvsSNR} that for \ac{snr} values larger than $8$ dB, Meta-ViterbiNet consistently achieves the lowest  \ac{ber} values among all considered data-driven receivers, with gains of up to 0.5dB over the online training counterpart.

\begin{figure*}
    \centering
    \begin{subfigure}[b]{0.48\textwidth}
    \includegraphics[width=\textwidth,height=0.22\textheight]{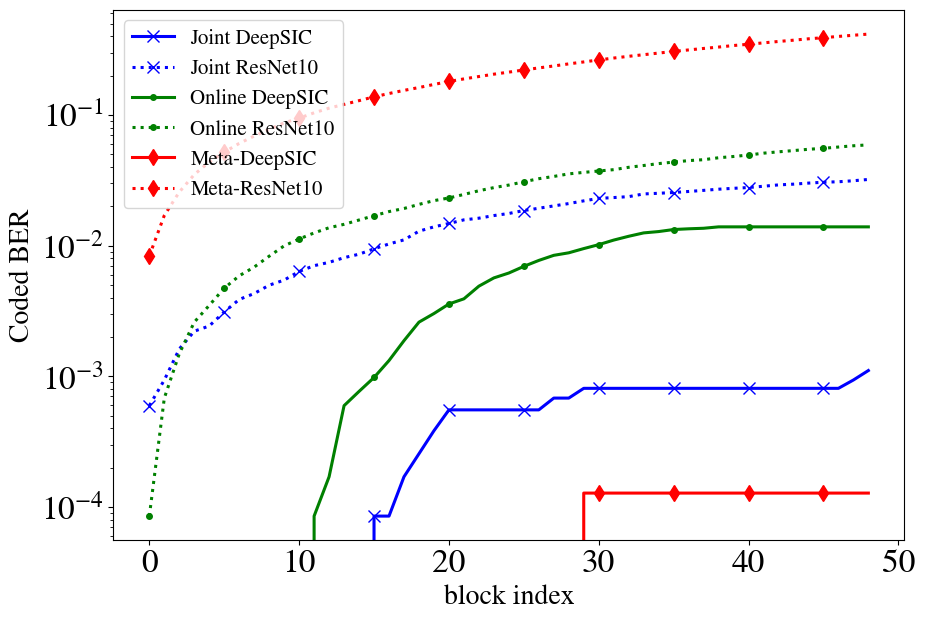}
    \caption{Coded \ac{ber} vs. block index, ${\rm SNR} = 14$ dB.}
    \label{fig:SyntheticBERvsBlockMIMO}
    \end{subfigure}
    \begin{subfigure}[b]{0.48\textwidth}
    \includegraphics[width=\textwidth,height=0.22\textheight]{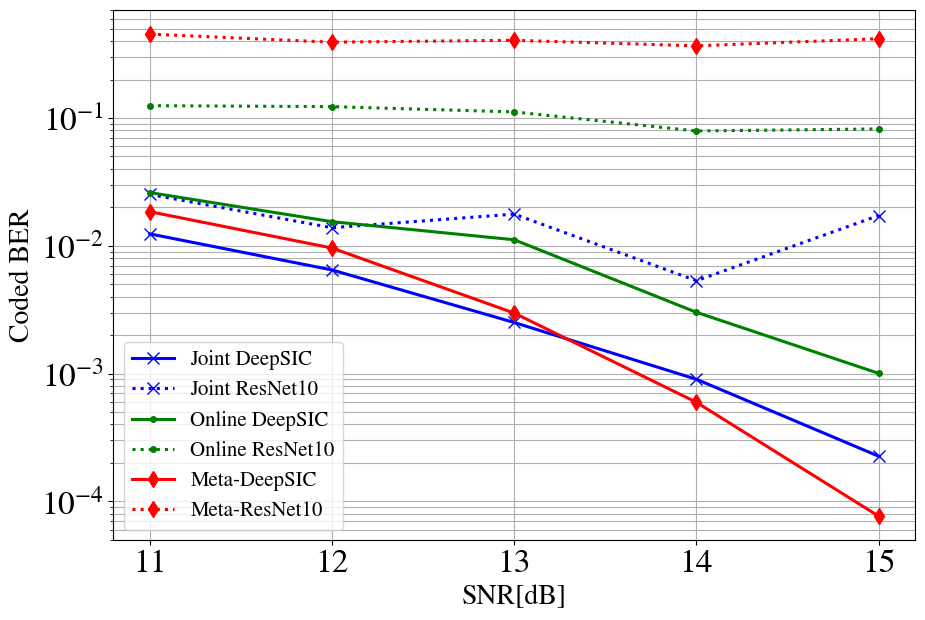}
    \caption{Coded \ac{ber} after 50 blocks vs. SNR.}
    \label{fig:SyntheticBERvsSNRMIMO}
    \end{subfigure}
    \caption{\ac{mimo} synthetic linear Gaussian channel, $\Blklen = 152$.}
    \label{fig:SyntheticBERMIMO} 
    \figSpace
\end{figure*}

\subsubsection{Memoryless MIMO Channels} 
For the \ac{mimo} setting, use $T_p=T_d=50$ blocks. The block length is set to $\Blklen = 152$ symbols, encoding 120 information bits with \acl{rs} [19,15] code. The input-output relationship of a the memoryless Gaussian \ac{mimo} channel is given by
\begin{equation}
\label{eqn:GaussianMIMO}
\myY = \myMat{H}\myS + \myVec{W},
\end{equation}
where $\myMat{H}$ is a known deterministic $N\times K$ channel matrix, and $\myVec{W}$ consists of $N$ i.i.d Gaussian \acp{rv}.  
We set the number of users and antennas to $N=K=4$. The channel matrix $\myMat{H}$ models spatial exponential decay, and its entries are given by
$\left( \myMat{H}\right)_{n,k} = e^{-|n-k|}$, for each $n \in \{1,\ldots, N\}$, $ k \in \{1,\ldots, K\}$.  The transmitted symbols are generated from a \ac{bpsk} constellation in a uniform i.i.d. manner, i.e., $\mathcal{S}=\{\pm 1\}$.

The numerical results for the memoryless Gaussian channel \eqref{eqn:GaussianMIMO} are depicted in Fig.~\ref{fig:SyntheticBERMIMO}.
Note that Fig.~\ref{fig:SyntheticBERvsBlockMIMO} shows the average coded \ac{ber} of the receivers for ${\rm SNR} = 14$ dB for a single trial, while Fig.~\ref{fig:SyntheticBERvsSNRMIMO} shows a sweep over a range of \ac{snr} values after 50 blocks, averaged over 10 trials. Similarly to Meta-ViterbiNet in the finite-memory \ac{siso} setting, Meta-DeepSIC outperforms all other training methods in medium to high SNRs. Here, the black-box architectures perform quite poorly, unable to compete with the model-based deep architecture of DeepSIC due to short blocklengths and the highly limited volumes of data used for training. Training DeepSIC with Algorithm~\ref{alg:online-meta-learning} outperforms the joint training approach with gains of up to 1dB, and with gains of 1.5dB over online training, which notably struggles in tracking such rapidly time-varying and challenging conditions.

\vspace{-0.4cm}
\subsection{Non-Linear Synthetic Channel Results}
\label{subsec:syntNL_simulation_results}
\vspace{-0.1cm}
Next, we evaluate the proposed online meta-learning scheme in non-linear synthetic channels, where one can greatly benefit from the model-agnostic nature of \ac{dnn}-based receivers.

\begin{figure*}[t!]
    \centering
    \begin{subfigure}[b]{0.48\textwidth}
    \includegraphics[width=\textwidth,height=0.22\textheight]{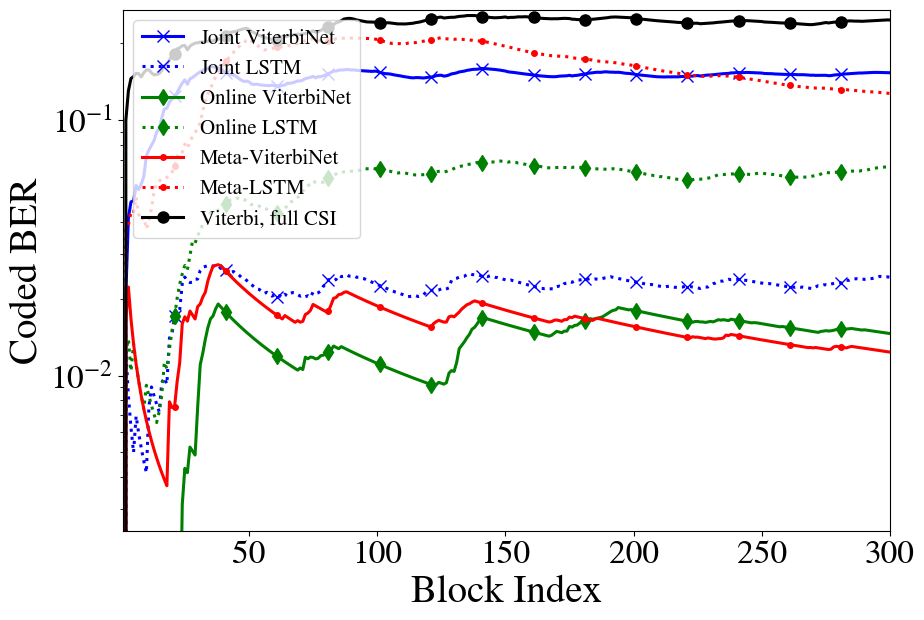}
    \caption{Coded \ac{ber} vs. block index, ${\rm SNR} = 12$ dB.}
    \label{fig:NonLinearSyntheticBERvsBlockSISO}
    \end{subfigure}
    \begin{subfigure}[b]{0.48\textwidth}
    \includegraphics[width=\textwidth,height=0.22\textheight]{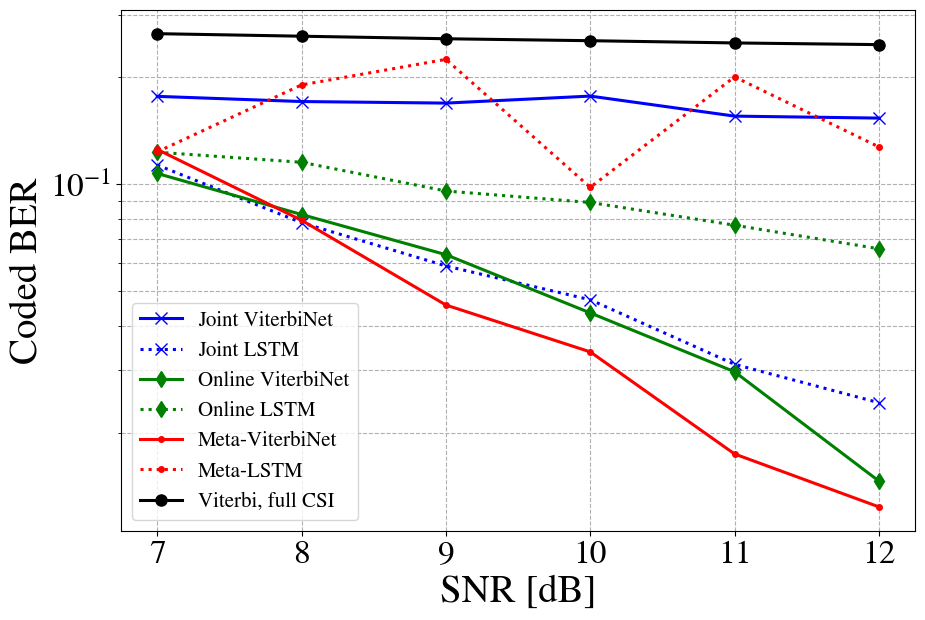}
    \caption{Coded \ac{ber} after 300 blocks vs. SNR.}
    \label{fig:NonLinearSyntheticBERvsSNRSISO}
    \end{subfigure}
    \caption{\ac{siso} synthetic non-linear Gaussian channel, $\Blklen = 136$.}
    \label{fig:NonLinearSyntheticBERSISO} 
    \figSpace
\end{figure*}

\subsubsection{\ac{siso} Finite-Memory Channels} To check the robustness of our method in a non-linear case, we generate a non-linear channel by applying a non-linear transformation to synthetic channel in \eqref{eqn:Gaussian}. The resulting finite-memory \ac{siso} channel is give by
\begin{equation}
\label{eqn:Gaussian1}
\myY_{i,j} =\tanh{\Big(C \cdot \Big(\sum_{l=0}^{\Mem-1} h_{l,j}S_{i-l,j} + w_{i,j}\Big)\Big)}.
\end{equation}
This operation may represent, e.g., non-linearities induced by the receiver acquisition hardware. The hyperparameter $C$ stands for a power attenuation at the receiver, and chosen empirically as $C=\frac{1}{2}$. All other hyperparameters and settings are the same as those used in the previous subsection. This simulation shows a consistent gain of around $0.75$dB over the SNRs with values of $8$dB - $12$dB, as observed in Fig.~\ref{fig:NonLinearSyntheticBERSISO}.

\subsubsection{Memoryless \ac{mimo} Channels} 
\begin{figure*}[t!]
    \centering
    \begin{subfigure}[b]{0.48\textwidth}
    \includegraphics[width=\textwidth,height=0.22\textheight]{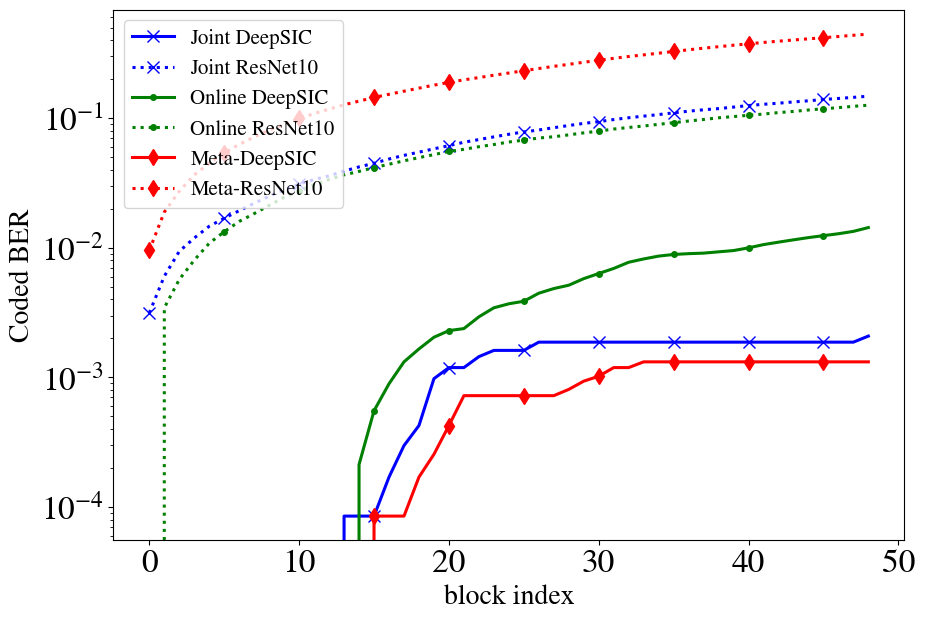}
    \caption{Coded \ac{ber} vs. block index, ${\rm SNR} = 14$ dB.}
    \label{fig:NonLinearSyntheticBERvsBlockMIMO}
    \end{subfigure}
    \begin{subfigure}[b]{0.48\textwidth}
    \includegraphics[width=\textwidth,height=0.22\textheight]{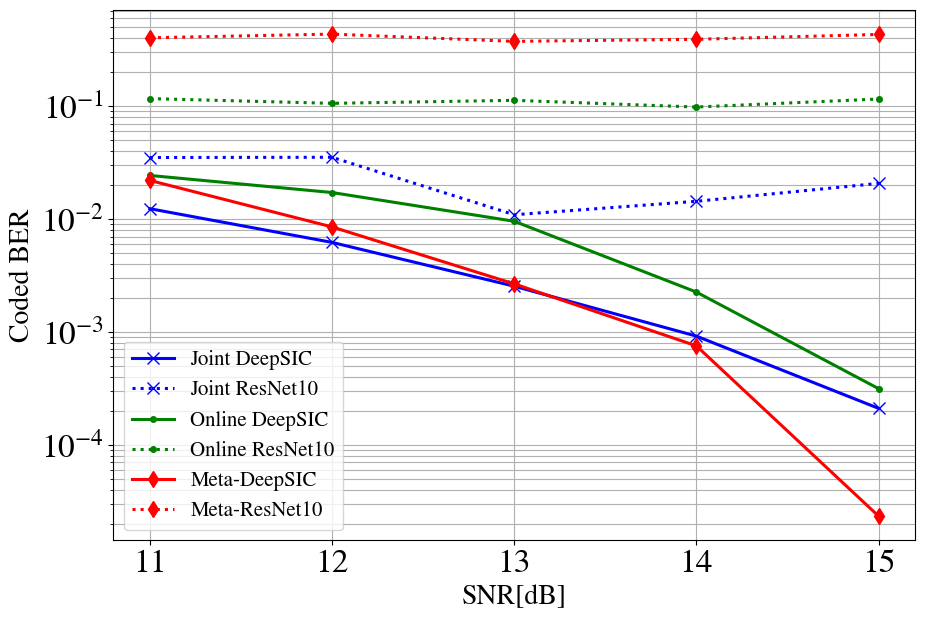}
    \caption{Coded \ac{ber} after 50 blocks vs. SNR.}
    \label{fig:NonLinearSyntheticBERvsSNRMIMO}
    \end{subfigure}
    \caption{\ac{mimo} synthetic non-linear Gaussian channel, $\Blklen = 152$.}
    \label{fig:NonLinearSyntheticBERMIMO} 
    \figSpace
\end{figure*}

Similarly to the \ac{siso} case, we simulate a non-linear \ac{mimo} channel from \eqref{eqn:GaussianMIMO} via
\begin{equation}
\label{eqn:Gaussian2}
\myY = \tanh{\Big(C \cdot\Big(\myMat{H}\myS + \myVec{W}\Big)\Big)},
\end{equation}
with $C=\frac{1}{2}$ and the remaining simulation settings are chosen as in Subsection~\ref{subsec:synth_simulation_results}. The numerical results for this channel are illustrated in Fig.~\ref{fig:NonLinearSyntheticBERMIMO}, showing that the superiority of our approach is maintained in non-linear setups. The right part shows an average over 20 trials, while the left graph corresponds to an exemplary trial. In particular, we note that gains of up to 0.5dB in high SNRs are achieved compared with online training, while the gap from joint method is around 0.25dB. Here, our gain becomes apparent as the SNR increases.

%-----------------------------------
%	Simulation Results
%-----------------------------------
\vspace{-0.4cm}
\subsection{COST 2100 Channel Results}
\label{subsec:cost_simulation_results}
\vspace{-0.1cm}

Next, we consider channels generated using the COST 2100 geometry-based stochastic channel model \cite{liu2012cost}, which is widely used in indoor wireless communications.

% \begin{figure*}
%     \centering
%     \begin{subfigure}[b]{0.48\textwidth}
%     \includegraphics[width=\textwidth,height=0.22\textheight]{fig/test_8x8_cost_ant1.png}
%     \caption{Antenna 1 channel.}
%     \label{fig:cost_channel_antenna1}
%     \end{subfigure}
%     %
%     \begin{subfigure}[b]{0.48\textwidth}
%     \includegraphics[width=\textwidth,height=0.22\textheight]{fig/test_8x8_cost_ant2.png}
%     \caption{Antenna 2 channel.}
%     \label{fig:cost_channel_antenna2}
%     \end{subfigure}
%     %
%     \caption{\ac{mimo} COST 2100 channels.}
%     \label{fig:CostBER} 
%     \figSpace
% \end{figure*}

\subsubsection{\ac{siso} Finite-Memory Channels} 
\begin{figure*}
    \centering
    \begin{subfigure}[b]{0.48\textwidth}
    \includegraphics[width=\textwidth,height=0.22\textheight]{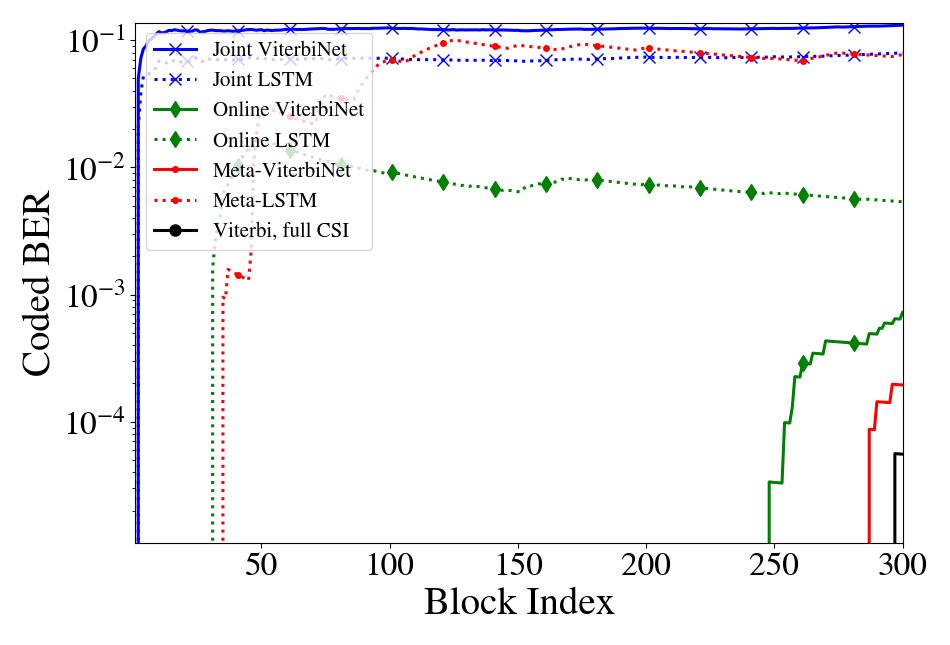}
    \caption{Coded \ac{ber} vs. block index, ${\rm SNR} = 12$ dB.}
    \label{fig:SISOcostBERvsBlock}
    \end{subfigure}
    \begin{subfigure}[b]{0.48\textwidth}
    \includegraphics[width=\textwidth,height=0.22\textheight]{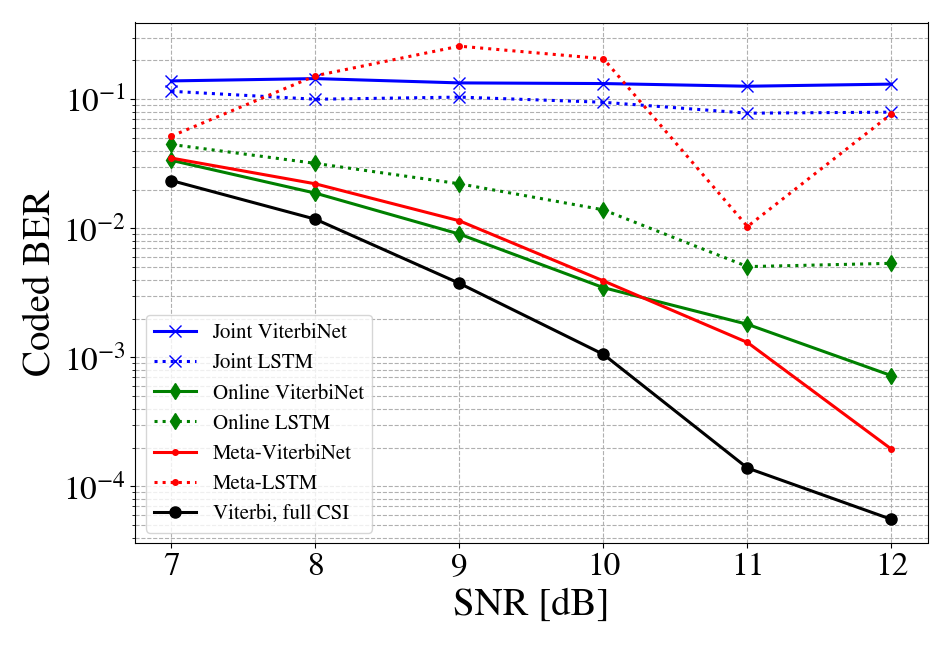}
    \caption{Coded \ac{ber} after 300 blocks vs. SNR.}
    \label{fig:SISOcostBERvsSNR}
    \end{subfigure}
    \caption{\ac{siso} COST 2100 channel, $\Blklen = 136$.}
    \label{fig:SISOCostBER} 
    \figSpace
\end{figure*}

We generate each realization of the taps using an indoor hall $5$ GHz setting with single-antenna elements. We use the same block length and number of error-correction symbols, as well as the same initial training set $\mySet{D}_0$ as in the synthetic model. This setting may represent a user moving in an indoor setup while switching between different microcells. Succeeding in this scenario requires high adaptivity since there is considerable mismatch between the train and test channels. The test is carried out using a sequence of difference realizations illustrated in Fig.~\ref{fig:COSTChannel}, whereas the initial training still follows the taps illustrated in Fig.~\ref{fig:synthetic_train_channel}. The two main figures, Fig.~\ref{fig:SISOcostBERvsBlock} and Fig.~\ref{fig:SISOcostBERvsSNR} illustrate gains of up to 0.6dB by using our scheme compared to other methods.

\subsubsection{Memoryless \ac{mimo} Channels} 
\begin{figure*}
    \centering
    \begin{subfigure}[b]{0.48\textwidth}
    \includegraphics[width=\textwidth,height=0.22\textheight]{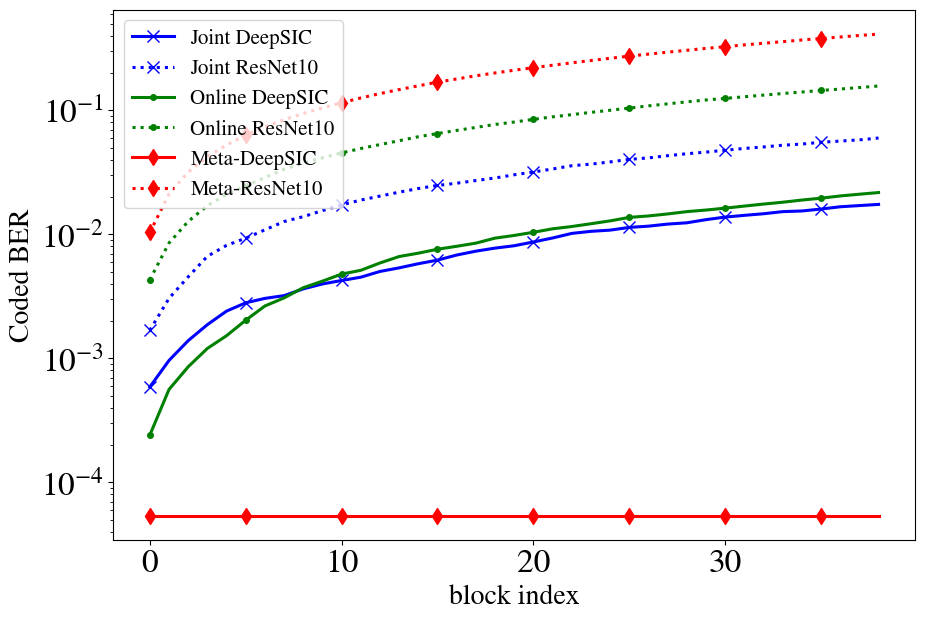}
    \caption{Coded \ac{ber} vs. block index, ${\rm SNR} = 14$ dB.}
    \label{fig:costBERvsBlockMIMO}
    \end{subfigure}
    \begin{subfigure}[b]{0.48\textwidth}
    \includegraphics[width=\textwidth,height=0.22\textheight]{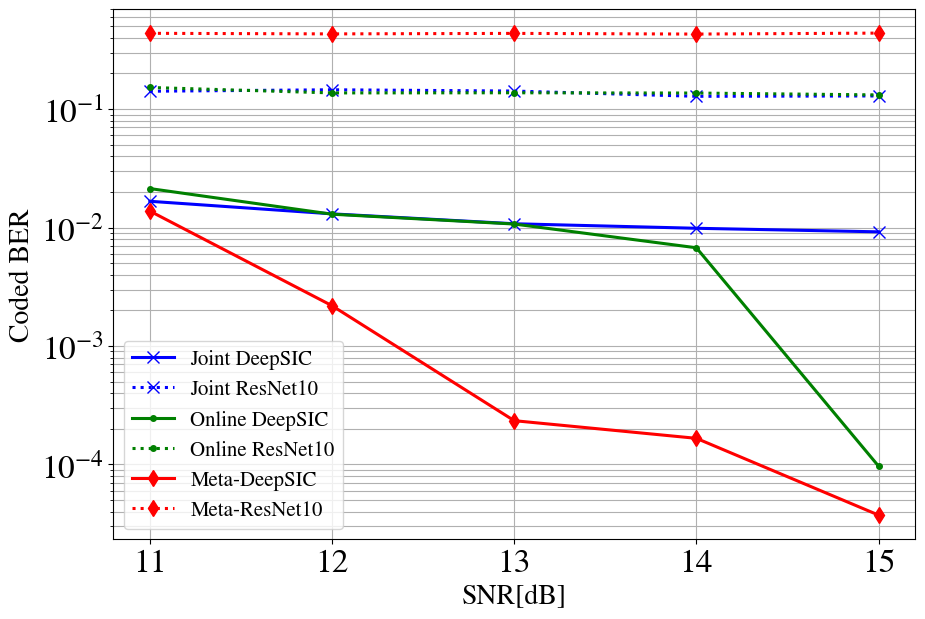}
    \caption{Coded \ac{ber} after 40 blocks vs. SNR.}
    \label{fig:costBERvsSNRMIMO}
    \end{subfigure}
    \caption{\ac{mimo} COST 2100 channel, $\Blklen = 152$.}
    \label{fig:CostBERMIMO} 
    \figSpace
\end{figure*}

For the current \ac{mimo} setting, we compose the test channels from only $T_p=T_d=40$ blocks and increase the number of users and antennas to $N = K = 8$. To simulate the channel matrix $\myMat{H}$, we follow the above \ac{siso} description and create $8 \times 8 = 64$ \ac{siso} COST 2100 channels; Each simulated channel corresponds to a single entry $\left( \myMat{H}\right)_{n,k}$ with $n \in \{1,\ldots, N\}$, $ k \in \{1,\ldots, K\}$. Moving to results, one may observe the respective graphs Fig.~\ref{fig:costBERvsBlockMIMO} and Fig.~\ref{fig:costBERvsSNRMIMO}, which show an unusual gain of up to 2dB approximately, as the joint and online methods are unable to compete with the proposed meta approach. The results here are averaged over 20 trials.

\begin{figure*}
    \centering
    \begin{subfigure}[b]{0.48\textwidth}
    \includegraphics[width=\textwidth,height=0.22\textheight]{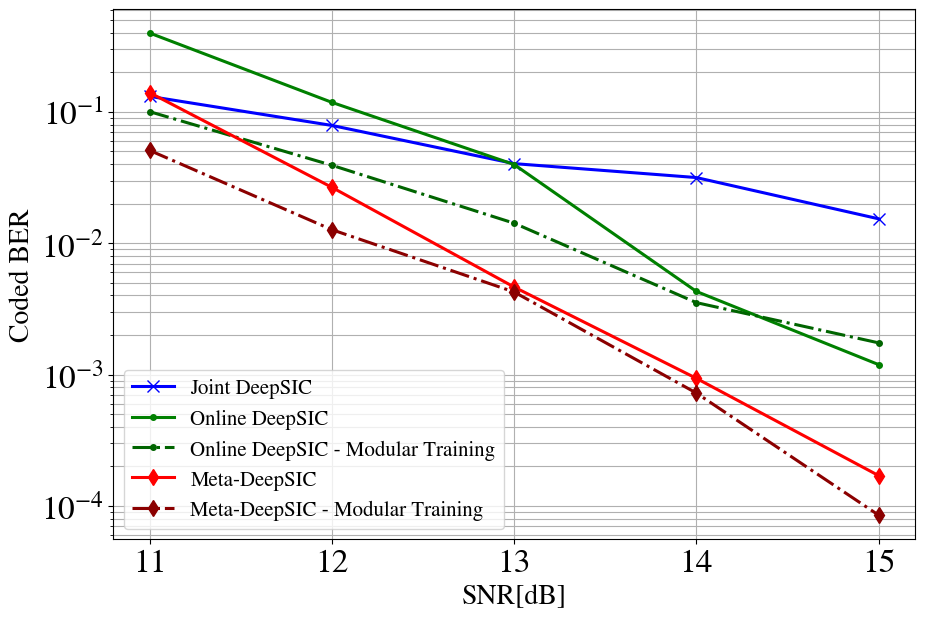}
    \caption{Synthetic linear channel.}
    \label{fig:synth_modular_online}
    \end{subfigure}
    \begin{subfigure}[b]{0.48\textwidth}
    \includegraphics[width=\textwidth,height=0.22\textheight]{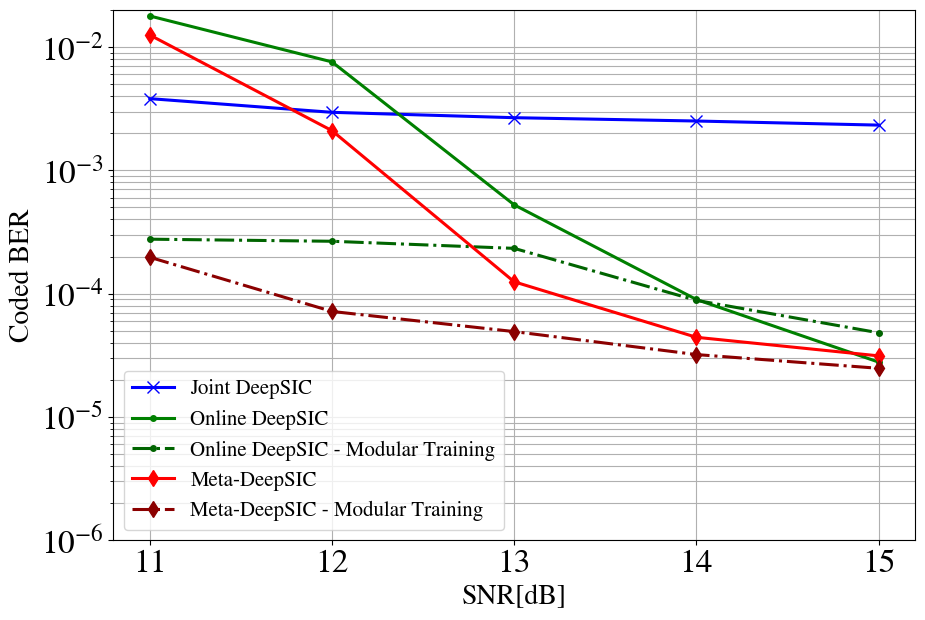}
    \caption{COST 2100 channel.}
    \label{fig:cost_modular_online}
    \end{subfigure}
    \caption{Modular Training - Coded \ac{ber} vs. \ac{snr}, $\Blklen = 152$.}
    \label{fig:ModularTrainingOnline} 
    \figSpace
\end{figure*}

\begin{figure*}[t!]
    \centering
    \begin{subfigure}[b]{0.48\textwidth}
    \includegraphics[width=\textwidth,height=0.22\textheight]{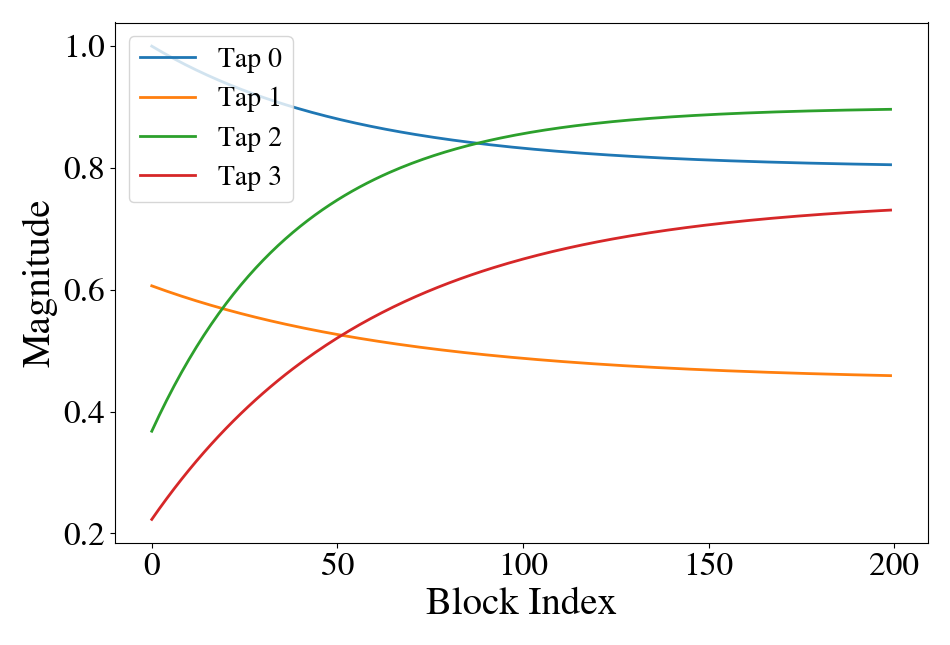}
    \caption{\ac{siso} synthetic non-periodic channel taps.}
    \label{fig:siso_non_periodic_channel_taps}
    \end{subfigure}
    \begin{subfigure}[b]{0.48\textwidth}
    \includegraphics[width=\textwidth,height=0.22\textheight]{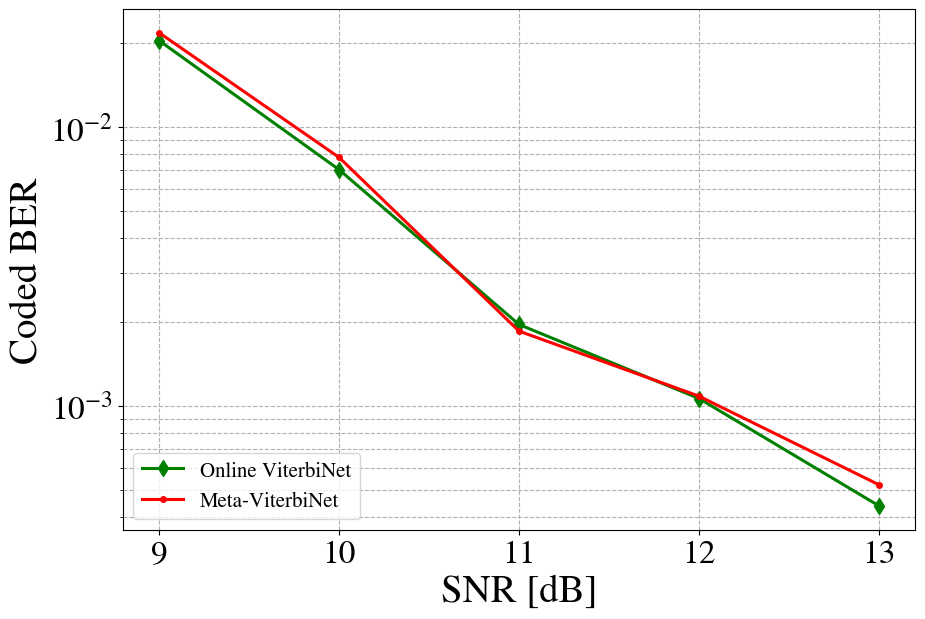}
    \caption{Coded \ac{ber} after 300 blocks vs. SNR.}
    \label{fig:siso_non_periodic_results}
    \end{subfigure}
    \caption{Exploring Non-Periodic Channel Profile, $\Blklen = 136$.}
    \label{fig:siso_non_periodic} 
    \figSpace
\end{figure*}

\subsection{Channel with non-Repetitive Temporal Variations} \label{subsec:nonperiodic}

The temporal variation profiles considered so far, illustrated in Fig.~\ref{fig:channels}, all exhibit periodic temporal variations, or more general patterns of variation of the channel. To elaborate on the importance of such patterns in enabling the benefits of meta-learning, we now simulate a \ac{siso} linear Gaussian channel, as in Subsection~\ref{subsec:synth_simulation_results} under the channel taps variations profile depicted in Fig.~\ref{fig:siso_non_periodic_channel_taps}. The channel variations are unstructured, limiting the ability of meta-learning to predict the variation profiles from  past channel realizations.
In Fig.~\ref{fig:siso_non_periodic_results}, we compare the ViterbiNet architecture trained online in a self-supervised manner both with and without predictive meta-learning. The figure demonstrates that predictive meta-learning is indeed beneficial when the temporal variations exhibit useful structure across blocks, enabling meta-generalization \cite{jose2020information}.

\subsection{Modular Training Results}
\label{subsec:modular-training}
To evaluate the modular training methods proposed in Section~\ref{sec:modular-training}, we next consider a multi-user \ac{mimo} setting. Here, the temporal variations stem from the fact that only a single user is changing ($k'=2$), with the other ones being static. The receiver uses DeepSIC, and knows the identity of the dynamic user. We numerically compare the gains of exploiting this knowledge using Algorithm~\ref{alg:modular-training}, as well as the individual gains of combining modular training with predictive meta-learning compared with online adapting solely the network weights.

% We apply the modular training method: Training the neural networks of a single given user, as described in Subsection~\ref{subsec:modular_training_formulation} and Subsection~\ref{subsec:modular_training_example}.
Fig.~\ref{fig:ModularTrainingOnline} shows the two compared methods -- online-training and (predictive) online-meta training -- executed either with or without modular training (Algorithm~\ref{alg:modular-training}) for the synthetic linear channel (Fig.~\ref{fig:synth_modular_online}) and the COST 2100 channel (Fig.~\ref{fig:cost_modular_online}). We also plot the joint training method for completeness. 
One may observe that our modular meta-training achieves superior results compared to the other methods. As depicted in Fig.~\ref{fig:synth_modular_online}, one may gain additional 0.3 dB under the meta-learning scheme by employing the modular training approach, for low and high \ac{snr} values. In medium \ac{snr}, one is expected to achieve similar results to the non-modular method. On the other hand, Fig.~\ref{fig:cost_modular_online} shows that the combined approach can provide substantial benefits, due to the rapid variations of the channel. Specifically, the modular gains appear in low to medium \ac{snr}, while the gains of the meta-learning stage appear for the higher SNR values. The gain in both scenarios for the two-stage approach versus simple online training is up to 2.5 dB. These results indicate the gains of our two-stage approach, and show that Algorithm~\ref{alg:modular-training} translates the interpretable modular architecture of DeepSIC into improved online adaptation in the presence of temporal variations due to a combination of mobile and static users.

\subsection{Complexity Analysis}
\label{subsec:complexity-analysis}
We now compare the complexity of the different training methods in terms of the number of iterations. 
Joint training does not apply re-training, and hence it has the lowest possible complexity at test time. For self-supervised online training, re-training is done for $I_{\rm sgd}$ SGD iterations at every coherence interval. The proposed predictive meta-learning scheme performs online training at each coherence interval with $I_{\rm sgd}$ iterations. Furthermore, it optimizes the initial weights via $I_{\rm meta}$ learning iterations, which are repeated periodically every $F$ blocks. 
Consequently, meta-learning applies on average $\frac{I_{\rm sgd} + I_{\rm meta}}{F}$ iterations on each coherence duration.

Nonetheless, the proposed meta-learning update, applied once each $F$ block, can help online training, applied on each block, to converge more quickly. As a result, while the proposed approach comes at the cost of $I_{\rm meta}/F$ additional gradient computations per block on average, it can operate with fewer online training iterations, $I_{\rm sgd}$. Therefore, meta-learning can in fact reduce the overall number of gradient steps as compared with online training.

\subsection{Meta-Learning Frequency Effect} \label{subsec:MetaFreq}

\begin{figure*}[t!]
    \centering
    \begin{subfigure}[b]{0.48\textwidth}
    \includegraphics[width=\textwidth,height=0.22\textheight]{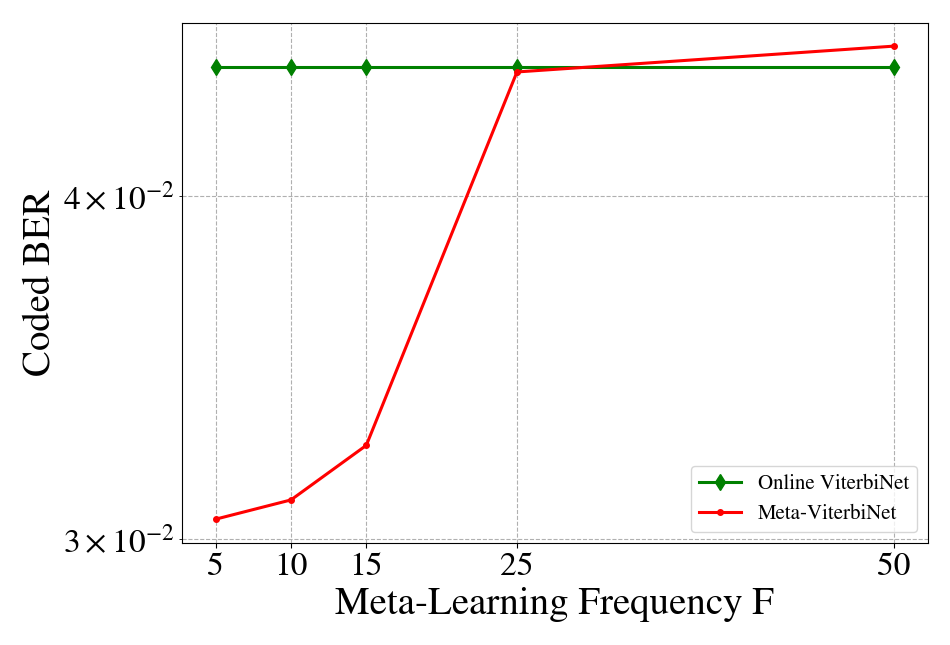}
    \caption{\ac{siso} synthetic linear Gaussian channel.}
    \label{fig:synthetic_non_linear_meta_learning_frequency}
    \end{subfigure}
    \begin{subfigure}[b]{0.48\textwidth}
    \includegraphics[width=\textwidth,height=0.22\textheight]{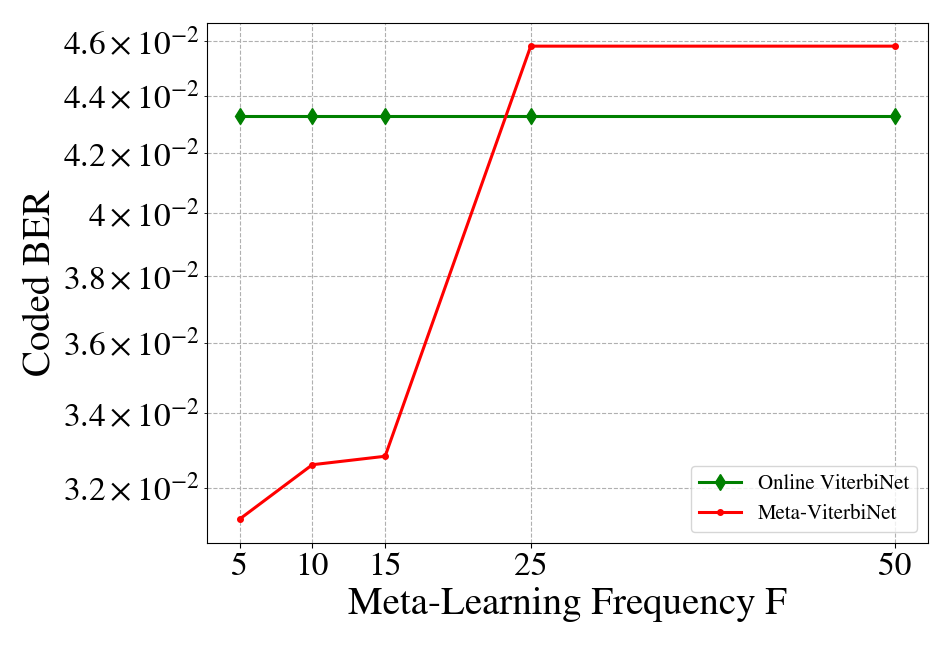}
    \caption{\ac{siso} synthetic non-linear Gaussian channel.}
    \label{fig:synthetic_non_linear_meta_learning_frequencyB}
    \end{subfigure}
    \caption{Coded \ac{ber} versus meta-learning frequency $F$; ${\rm SNR}=10$ dB, $\Blklen = 136$.}
    \label{fig:meta_learning_frequency} 
    \figSpace
\end{figure*}

We conclude our numerical evaluations by  studying the effect of changing the meta-learning frequency $F$. To this end, fixing ${\rm SNR} = 10$ dB, $\Blklen = 136$, as in Subsection~\ref{subsec:synth_simulation_results} and Subsection~\ref{subsec:syntNL_simulation_results}, Fig.~\ref{fig:meta_learning_frequency} illustrates the coded \ac{ber} after 1000 blocks, plotted against values of $F \in \{5,10,15,25,50\}$. We observe that a lower value of $F$ yields a lower \ac{ber} for both the linear and non-linear cases.  It is also observed that increasing the value of $F$ causes the performance of meta-learning to be within a minor gap as that of the online training. 
The small difference follows since, while for large values of $F$, meta-learning  is rarely carried out, the initial weights utilized at each block differ from that of conventional online training. In particular, when meta-learning is employed, the same initial weights are used for consecutive $F$ blocks, while online training sets the initial weights at block $j$ to be those utilized  in  block $j-1$.

	\section{Conclusion}
	\label{sec:Conclusions}
	\vspace{-0.1cm}
In this paper, we proposed a two-stage training method whose goal is to aid \ac{dnn}-based receivers in tracking time-varying channels. The first part of the paper introduces a predictive meta-learning method that incorporates both short-term and long-term relations between the symbols and the received channel values. This method yields initial weights such that training on a block transmitted over some channel minimizes the error on the block transmitted over the next channel. The approach is generic and applicable to any \ac{dnn}-based receiver. Then, we have introduced a modular training scheme that exploits the interpretable structure of model-based deep receivers, and applies meta-learning to allow efficient adaptation only for the subset of modules that suffers mostly from rapidly varying channels. %to further reduce the overall transmission error}. 
Numerical studies demonstrate that, by properly integrating these methods, model-based deep receivers trained with the meta-training algorithm and modular training  outperform existing self-supervision and joint learning approaches \cite{oshea2017introduction, xia2020note,shlezinger2019viterbinet, teng2020syndrome}.
	%----------------------------------------------------------------------------------------
	%	BIBLIOGRAPHY
	%----------------------------------------------------------------------------------------
	%\vspace{-0.2cm}
	\bibliographystyle{IEEEtran}
	\bibliography{IEEEabrv,ms}
	
\end{document}